\documentclass[12pt]{article}

\pdfoutput=1
\usepackage{color,amsmath,amssymb,exscale,psfrag,epsfig}
\usepackage{cite,color,url}
\usepackage{pdfpages}
\usepackage{graphicx}
\usepackage{slashed}
\usepackage[T1]{fontenc}
\usepackage{xcolor}
\usepackage{accents}
\usepackage[colorlinks=true
,urlcolor=blue
,anchorcolor=blue
,citecolor=blue
,filecolor=blue
,linkcolor=blue
,menucolor=blue
,linktocpage=true
,pdfproducer=medialab
]{hyperref}
\input epsf
\def\bea{\begin{eqnarray}}
\def\eea{\end{eqnarray}}

\definecolor{nicered}{rgb}{0.7,0.1,0.1}
\definecolor{nicegreen}{rgb}{0.1,0.5,0.1}
\hypersetup{colorlinks,citecolor= nicegreen,linkcolor= nicered}

\newcommand{\tttt}{t\bar{t}t\bar{t}}

\catcode`\@=11
\def\lsim{\mathrel{\mathpalette\@versim<}}
\def\gsim{\mathrel{\mathpalette\@versim>}}
\def\@versim#1#2{\vcenter{\offinterlineskip
\ialign{$\m@th#1\hfil##\hfil$\crcr#2\crcr\sim\crcr } }}
\catcode`\@=12
\catcode`@=12
\begin{document}
\thispagestyle{empty}
\begin{flushright}
ICAS 025/16
\end{flushright}
\vspace{0.3in}
\begin{center}
{\Large \bf Four Tops for LHC} \\
\vspace{0.5in}
{\bf Ezequiel Alvarez$^{(a)\dagger}$,
Darius A. Faroughy$^{(b)\ddag}$,\\[1ex]
Jernej F. Kamenik$^{(b,c)\star}$,
Roberto Morales$^{(d)\ast}$,
Alejandro Szynkman$^{(d)\diamond}$
}
\vspace{0.2in} \\
{\sl $^{(a)}$ International Center for Advanced Studies (ICAS), UNSAM, Campus Miguelete\\
25 de Mayo y Francia, (1650) Buenos Aires, Argentina }
\\[1ex]
{\sl $^{(b)}$ Jo\v zef Stefan Institute\\ Jamova 39, 1000 Ljubljana, Slovenia 
}
\\[1ex]
{\sl $^{(c)}$ Faculty of Mathematics and Physics, University of Ljubljana\\
Jadranska 19, 1000 Ljubljana, Slovenia
}
\\[1ex]
{\sl $^{(d)}$ IFLP, Dpto. de F\'isica, CONICET, UNLP \\ C.C. 67, 1900 La Plata, Argentina
}
\end{center}
\vspace{0.5in}

\begin{abstract}
We design a search strategy for the Standard Model $\tttt$ production at the LHC in the same-sign dilepton and trilepton channels.  We study different signal features and, given the small expected number of signal events, we scrutinize in detail all reducible and irreducible backgrounds. Our analysis shows that by imposing a basic set of jet and lepton selection criteria, the SM $pp \to \tttt$ process could be evidenced in the near future, within Run-II, when combining both multi-lepton search channels. We argue that this search strategy should also be used as a guideline to test New Physics coupling predominantly to top-quarks. In particular, we show that a non-resonant New Physics enhancement in the four-top final state would be detectable through this search strategy. We study two {\it top-philic} simplified models of this kind, a neutral scalar boson and a $Z^\prime$, and present current and future exclusion limits on their mass and couplings.
\end{abstract}

\vspace*{20mm}
\noindent {\footnotesize E-mail:
{\tt 
$\dagger$ \href{mailto:sequi@df.uba.ar}{sequi@df.uba.ar},
$\ddag$ \href{mailto:darius.faroughy@ijs.si}{darius.faroughy@ijs.si},
$\star$ \href{mailto:jernej.kamenik@ijs.si}{jernej.kamenik@ijs.si},\\
$\ast$ \href{mailto:roberto.morales@fisica.unlp.edu.ar}{roberto.morales@fisica.unlp.edu.ar},
$\diamond$ \href{mailto:szynkman@fisica.unlp.edu.ar}{szynkman@fisica.unlp.edu.ar}
}}


\newpage


\section{Introduction}
\label{section:1}

With the discovery of the last elementary particle predicted by the Standard Model (SM), the Higgs boson, and with the recent upgrade in both energy and luminosity of the LHC, the field of high energy physics has entered unchartered waters. Concerning the discovery of New Physics (NP) at the LHC, we currently find ourselves at an inflection point with two possible outcomes: on one hand, NP may soon appear during Run-II in one of the signatures already being explored at the LHC. On the other hand, NP might as well be hidden in challenging signatures difficult to measure at current luminosities. In our experimental quest towards discovering NP at the LHC, it is therefore important to scrutinize all possible final states, including those produced in rare SM processes with small cross-sections, and usually considered to be beyond the LHC Run-II sensitivity reach.

Most current studies of this kind are geared towards final states and processes involving the heaviest known particles, the Higgs boson and the top quark. In particular, many recent proposals exist to study the Higgs boson in final states with extra radiation~\cite{ATLAS:2016lgh}, differential distributions in Higgs production~\cite{Aad:2015tna,Aad:2015lha,Khachatryan:2015yvw,Khachatryan:2015rxa}, Higgs pair production~\cite{CMS:2016cdj,Aaboud:2016xco,Khachatryan:2016sey} and Higgs production in association with other massive particles~\cite{ATLAS:2016pkl} including top-quarks~\cite{CMS:2016ygt,CMS:2016vqb,ATLAS:2016axz}.  On the other hand, the top quark has been studied extensively in LHC mainly through top-quark pair production~\cite{ATLAS:2016soq,CMS:2016xyh,Aaboud:2016iot,CMS:2016cue}, as well as through single-top production~\cite{ATLAS:2016lte,CMS:2016ufa}, both within the SM and beyond. More recently, associate $t\bar t H$ and $t\bar t Z$ production have also become objects of intense study~\cite{ATLAS-CONF-2016-003,CMS:2016dui}.  Finally, one of the long standing challenges in top physics is to measure four-top quark production. This process is interesting given that its small cross-section in the SM can be significantly enhanced in many NP scenarios, see Ref.~\cite{Lillie:2007hd,Pomarol:2008bh,Kumar:2009vs,Cacciapaglia:2011kz,Perelstein:2011ez,AguilarSaavedra:2011ck,Beck:2015cga,Dev:2014yca}. In this study we propose a dedicated search strategy for the SM production of four-top quarks at the 13 TeV LHC in the multi-lepton decay channels. We probe the reach of our search and show that it is possible for the LHC to find evidence for four-top production before the end of Run-II.

\begin{figure}[!t]
\begin{center}
\includegraphics[width=0.5\textwidth]{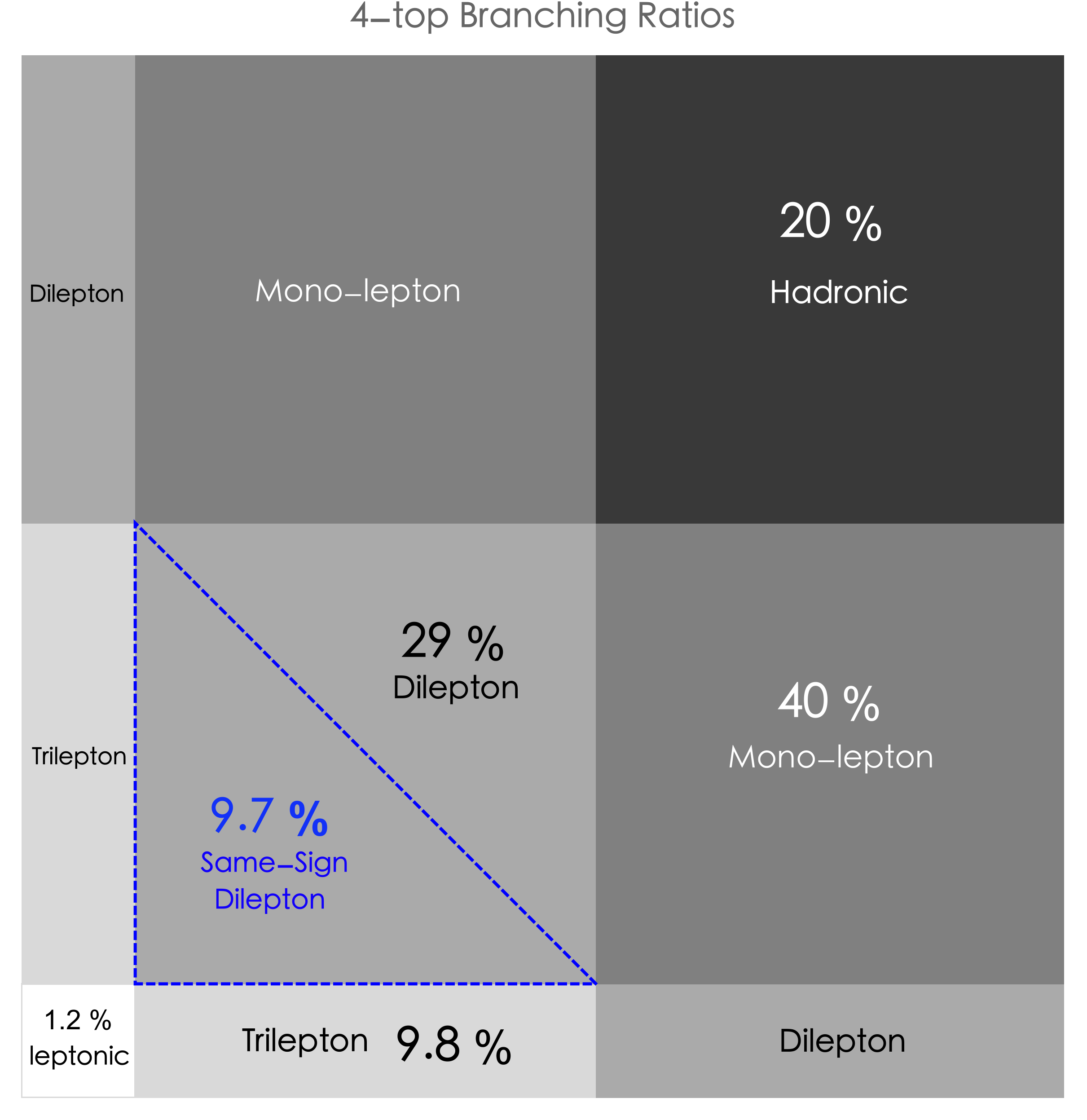}~
\caption{
\small Branching ratios for the possible $\tttt$ decay channels. The total square has unit area and each side represents either the leptonic, semi-leptonic or hadronic branching ratio of  $t\bar t$ (undecayed $\tau$-lepton is included). For the sake of clarity we have rendered all areas corresponding to a particular final state with the same color (even disconnected in the graphic), whereas the quoted percentage corresponds to the sum of all areas associated to a particular final state.
}
\label{BRsquare}
\end{center}
\end{figure}

The SM four-top process produces a rich set of final states all giving rise to interesting signatures at the LHC. A useful way of illustrating all four-top decay channels is depicted in Fig.~\ref{BRsquare}, where each side of the unit {\it square} is partitioned into the leptonic, semi-leptonic and hadronic branching ratio of $t\bar t$. Regions with the same shade represent a particular $\tttt$ decay mode and the total area of the shades gives the corresponding branching ratio. The dominant decay mode is the mono-leptonic channel with a branching ratio of $\sim40$\%, followed by the fully hadronic and opposite-sign (OS) dilepton modes with $\sim20$\% each, the same-sign (SS) dilepton (represented by the dotted blue contour) and the trilepton modes with $\sim10$\% each and finally the fully leptonic mode with $\sim1\%$. Because of large backgrounds in the mono-leptonic and hadronic channels, the SS dilepton channel is usually considered to be the most promising search channel for the SM four-top process at the LHC. Bellow we argue that, for higher luminosities within Run-II, a dedicated search based on trileptons can achieve comparable or even better sensitivities to the four-top signal than the SS dilepton channel.

Four-top production in the SM is challenging due to its small cross-section at the LHC, of about ~10 fb at 13 TeV~\cite{Bevilacqua:2012em}. There are only few existing proposals for beyond SM four-top searches~\cite{Lillie:2007hd, Kumar:2009vs, Gerbush:2007fe, Acharya:2009gb, Cacciapaglia:2011kz, Perelstein:2011ez, AguilarSaavedra:2011ck,Dev:2014yca,Gregoire:2011ka, Liu:2015hxi, Gori:2016zto,Kim:2016plm} and also some preliminary experimental reports~\cite{Khachatryan:2014sca, Aad:2015kqa, ATLAS:2016gqb, CMS:2016wig,ATLAS:2016sno,Aad:2016tuk,CMS:2016ahn} in this direction. 
We present a detailed classification of all relevant backgrounds for the four-top SS dilepton and trilepton channels. Special emphasis is given to the reducible backgrounds comprised of fake leptons and charge-flip. These backgrounds, which are always difficult to model in a phenomenological analysis, must be correctly estimated since they turn out to be important for giving reliable predictions.    

A measurement or an upper bound on the SM four-top process using our proposed search strategy can be directly used to constrain interesting non-resonant NP scenarios coupling dominantly to top-quarks. These {\it top-philic} NP scenarios with SM-like kinematics would easily avoid bounds from top-pair production and other searches and would most naturally manifest themselves as deviations in the total four-top production. To illustrate this, we present current and projected constrains on two top-philic simplified models (a neutral vector and a scalar mediator) entering four-top production, and show that our search strategy can cover important regions of parameter space.

The remainder of the article is structured as follows. In Section \ref{section:2} we describe the expected features of the SM four-top process at the LHC relevant for the multi-lepton channels. In Section \ref{section:3} we design dedicated searches for the SS dilepton and trilepton channels separately and then combine them statistically to present discovery/evidence luminosities as well as exclusion limits for the SM four-top signal strength. To be as general as possible, we also consider the impact on signal sensitivity when using different estimations of the reducible backgrounds (fake leptons and charge-flip). In Section \ref{section:5} we use our SM four-top strategy to give bounds on two top-philic non-resonant NP models that contribute to $pp\to\tttt$ at the LHC. We give a final discussion and present our main conclusions in Section \ref{section:6}.  A detailed analysis of all irreducible and reducible backgrounds can be found in the Appendices.


\section{Signal features}
\label{section:2}

\begin{figure}[b!]
\begin{center}
\includegraphics[width=1\textwidth]{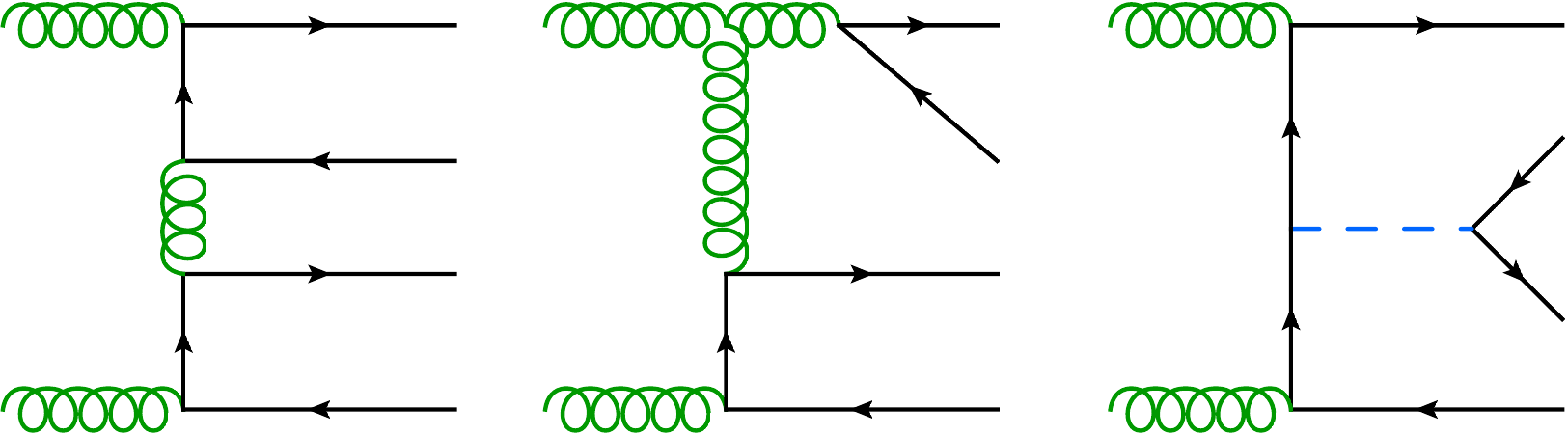}

\caption{
\small Gluon initiated representative LO diagrams contributing to $pp\to t \bar t t \bar t$ in the SM. The first two diagrams give leading contributions of order $\mathcal{O}(\alpha_S^4)$ while the last diagram includes a sub-leading contribution of order $\mathcal{O}(\alpha_S^2y_t^4)$ with a Higgs boson exchange (blue dashed propagator).
}
\label{diagrams}
\end{center}
\end{figure}

In the SM, the production of $\tttt$ is predominantly a QCD process of order $\mathcal{O}(\alpha_S^4)$ that requires a partonic center-of-mass energy of at least $4m_t \sim 692$ GeV, resulting in a very small cross-section at the LHC. Besides QCD, there is also a sub-leading Higgs boson exchange contribution of order $\mathcal{O}(\alpha_S^2y_t^4)$ and an EW contribution of order $\mathcal{O}(\alpha_S^2\alpha^2)$, both accounting for $\sim$10\% of the total cross-section. The QCD driven four-top process is given at leading order (LO) by 72 and 12 topologically inequivalent diagrams from initial $gg$ and $q\bar q$ scattering, respectively. At the LHC, the gluon-initiated process accounts for approximately $95$ \% of the total QCD cross-section. Using {\tt MadGraph5}~\cite{Alwall:2014hca} we calculated at $\sqrt s = 13$ TeV the four-top production cross-section and found $\sigma^{LO}(t \bar t t \bar t) = 9.7$ fb at LO and $\sigma^{NLO}(t \bar t t \bar t) = 12.32$ fb at next-to-leading order (NLO). For a complete analysis of $pp\to\tttt$ at NLO in QCD, the reader is referred to Ref.~\cite{Bevilacqua:2012em}. In Fig.~\ref{diagrams} we show  three representative Feynman diagrams contributing to the $gg\to \tttt$ process. 

\begin{figure}[t!]
\begin{center}
\includegraphics[width=0.48\textwidth]{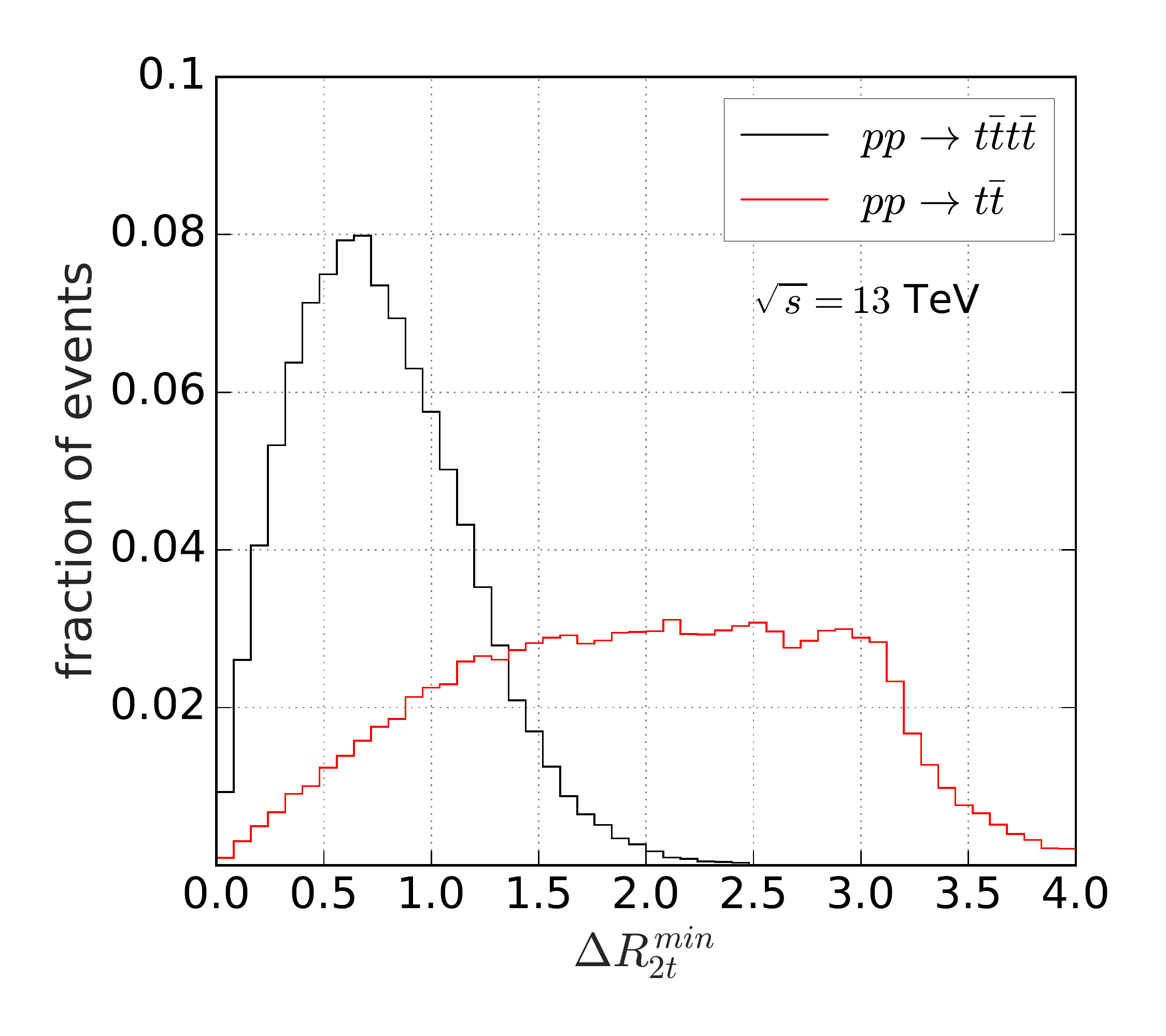}
\caption{
\small 
Minimum truth-level distance between any pair of top-quarks for both four-top production (black) and top-pair production (red).}
\label{angular}
\end{center}
\end{figure}
\begin{figure}[t!]
\begin{center}
\includegraphics[width=0.47\textwidth]{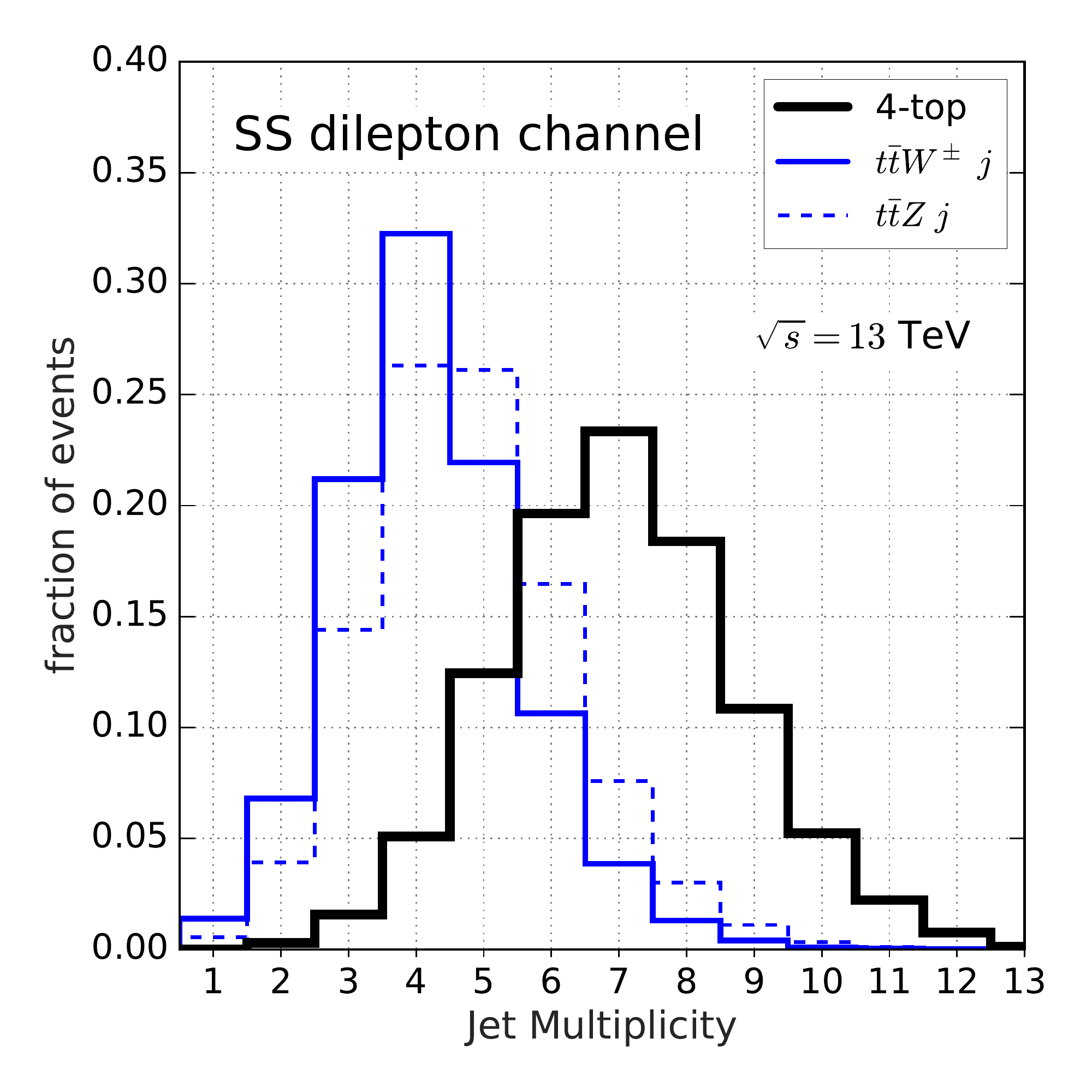}~
\includegraphics[width=0.47\textwidth]{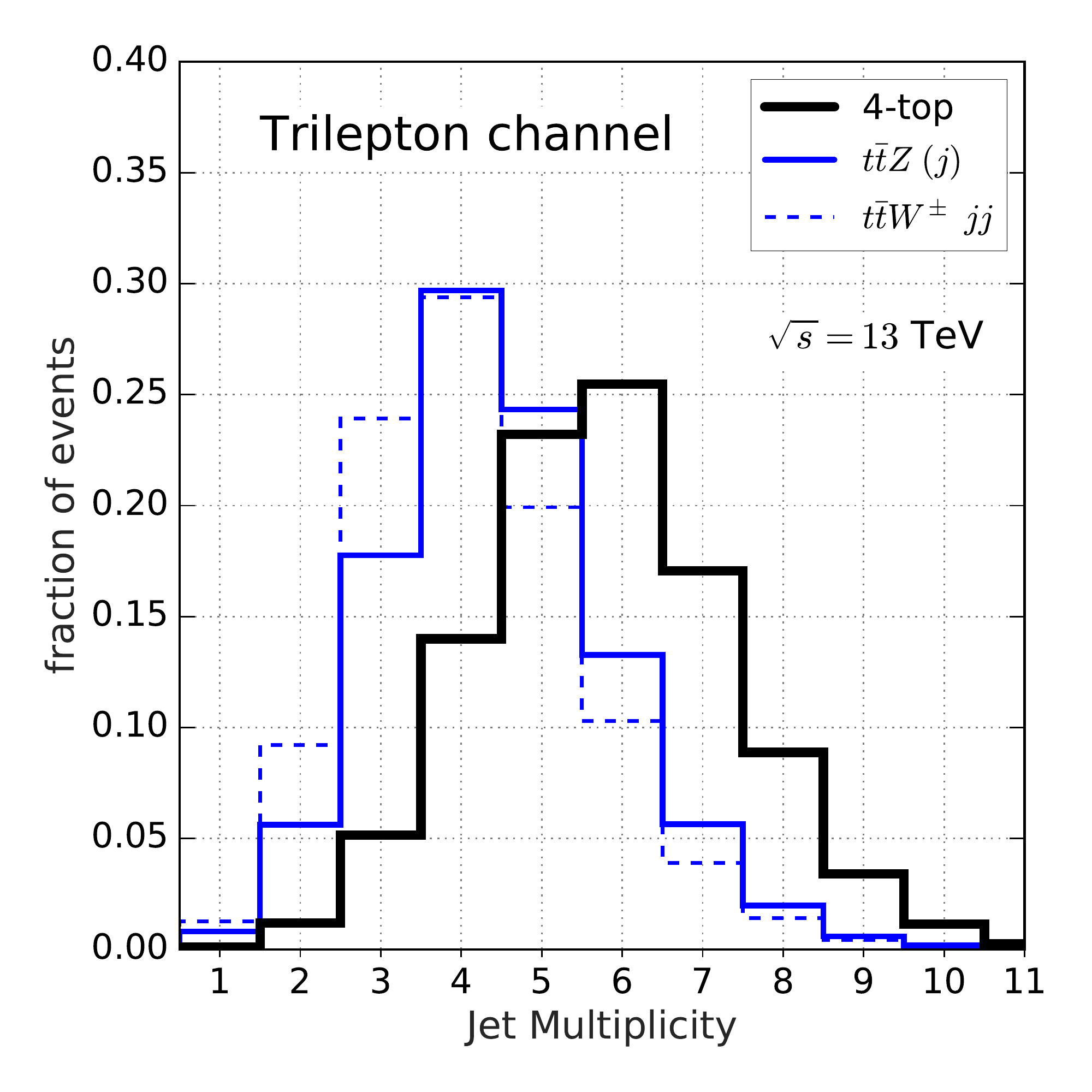}
\caption{
\small Distribution of the number of jets for the SS dilepton (left) and trilepton (right) channels for signal and expected backgrounds.
}
\label{njets}
\end{center}
\end{figure}

Since $\tttt$ decays into a large number of final states (of order $\mathcal{O}(10)$ particles) predominantly along the central direction, it is important to give a rough estimate of the amount of accidental object overlaps expected in the detector. For this we considered the minimum angular separation $\Delta R^{min}_{2t}$ between any pair of top-quarks in each event (irrespective of their charge). In Fig.~\ref{angular} we have plotted the truth-level distribution of this variable for the four-top signal (black), compared to top-pair production $pp\to t\bar t$ (red). The plot suggests that we should expect a considerable amount of object overlaps between the decay products of the four-top signal. This accidental overlap will manifest itself as a drop in isolation efficiency at the object reconstruction level, in particular it will show up as an increase in the fraction of leptons from top decays that end up accidentally close to jets. 

The main signature of a four-top event is the large number of $b$-jets coming from the weak decay of each top-quark. All of the dominant $t\bar t$ backgrounds are expected to have less $b$-jets per event, making the $b$-jet multiplicity $N_b$ the most important signal to background discriminant. In order to exploit this fact, we will use in our analysis a high efficiency operating point for the $b$-jet tagging algorithm and put a cut on the number of $b$-jets. This should be enough to raise  the signal-to-background ratio considerably.  

When focusing on the multi-lepton channels, besides having many b-jets, events are also expected to have a fair amount of hard light-quark and charm jets coming from the hadronic decay of top-quarks plus the expected additional soft jets from QCD radiation. For this reason the total jet multiplicity $N_j$ will also be a relevant variable in our search. In Fig.~\ref{njets} (left) we plot $N_j$ for the four-top sample in the SS dilepton channel. Here, the signal is characterized by a mean of $N_j=7$ hard jets, while the two leading backgrounds $t\bar t W$ and $t\bar t Z$ have jet multiplicity distributions peaking at lower values around $N_j=4$. Similar arguments hold for the trilepton channel, as can be seen in Fig.~\ref{njets} (right), the signal jet multiplicity peaks around $N_j=6$ also above the two leading backgrounds.

\begin{figure}[t!]
\begin{center}
\includegraphics[width=0.6\textwidth]{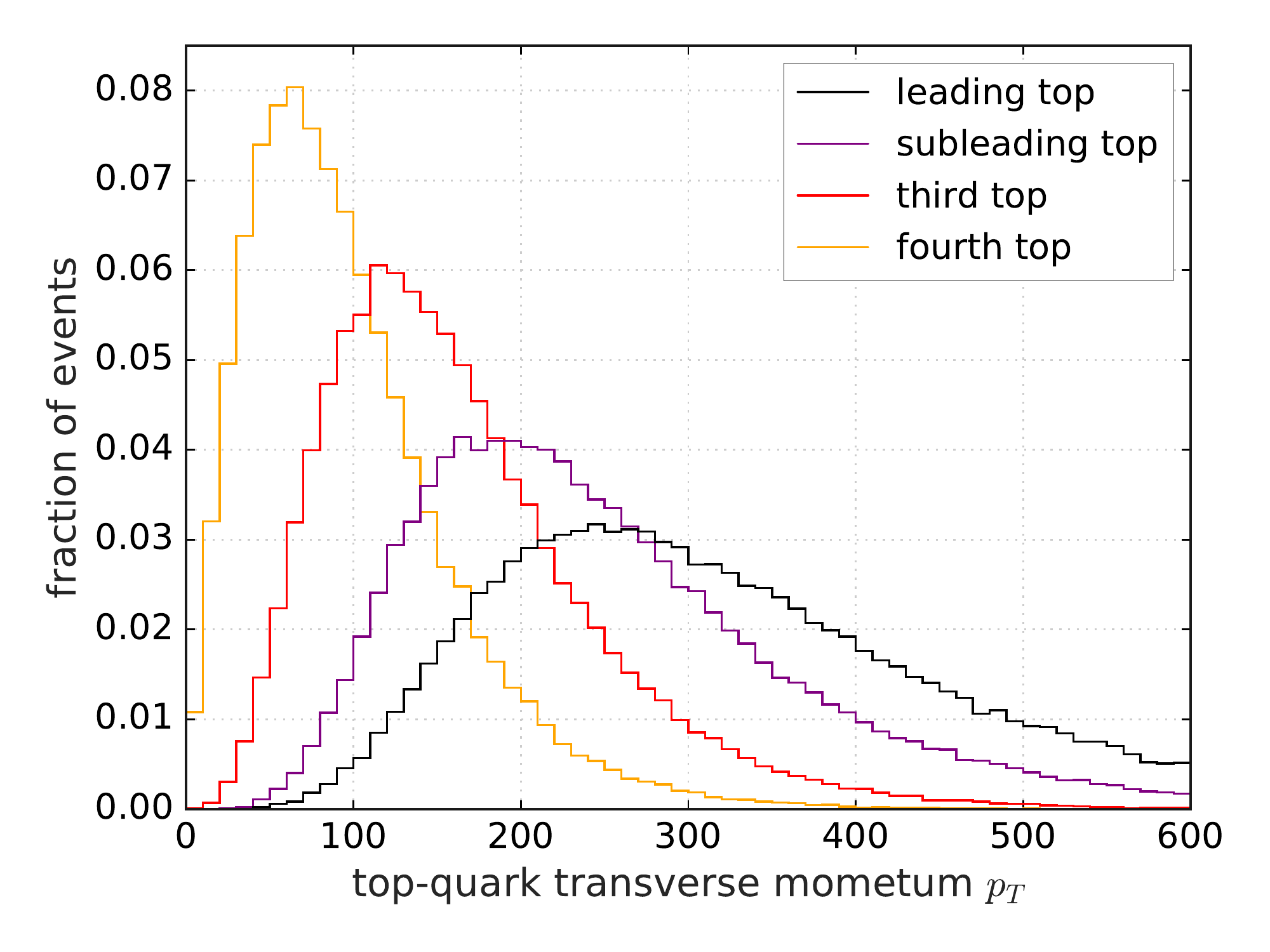}~
\caption{
\small $p_T$-distributions of  $p_T$-ordered top quarks in $\tttt$ production at 13TeV LHC. }
\label{pt-distribution}
\end{center}
\end{figure}

The remaining features characterizing the four-top signal in the multi-lepton channel are in part related to the total transverse energy of the process. Highly energetic events will have more boosted top-quarks giving rise to a very different signature when compared to events close to $\tttt$ production threshold. In order to get an idea of the distribution of events according to their transverse energy we show the truth level $p_T$-distribution for top-quarks in Fig.~\ref{pt-distribution}. 
We see that a significant fraction of events with boosted top-quarks (with $p_T \gtrsim 300$ GeV) can be expected. In fact, we find that 51\% of the events contain at least one top-quark with $p_T>300$ GeV, and 28\% (6\%) of the events contain at least two (three) top quarks with $p_T>300$ GeV.

Hadronic boosted top-quarks can be tagged using jet substructure techniques, see Ref.~\cite{Thaler:2008ju}. Unfortunately, at expected LHC luminosities and current top-tagging efficiencies, we find that even these sizeable boosted-top fractions are not sufficient and we do not include top-tagging in a competitive four-top search based on the multi-lepton signature. Such techniques might however be of more relevance in the single-lepton or fully hadronic $\tttt$ signatures and/or for larger signal event samples expected from the high-luminosity LHC phase.

\begin{figure}[t!]
\begin{center}
\includegraphics[width=0.85\textwidth]{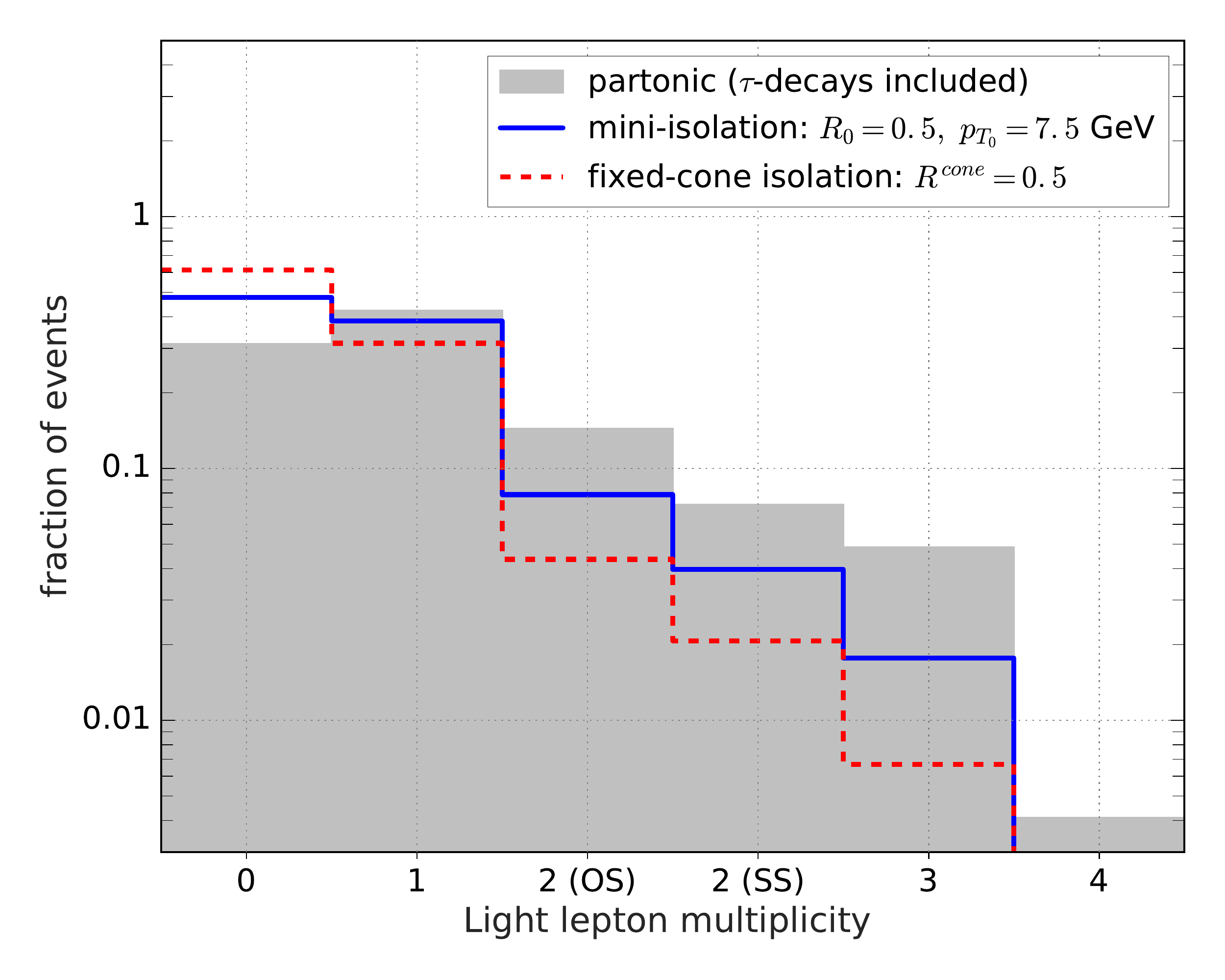}
\caption{
\small Four-top lepton multiplicity after taking into account detector effects and leptonic isolation criteria: mini-isolation (solid blue), traditional isolation criteria with cone radius of 0.5 (dashed red).
}
\label{leptons}
\end{center}
\end{figure}

Another important consequence of boosted top-quarks is however the reduction of the signal efficiency in the multi-lepton channels. The sought lepton coming from a boosted top-quark decay will usually be collimated with the $b$-quark and end up close to the $b$-jet axis, eventually overlapping with the b-jet induced hadronic activity in the calorimeters. Such signal leptons will fail the standard lepton isolation criteria. To bypass this issue as well as accidental lepton overlaps with other jets in the event, we use the mini-isolation technique, useful for identifying signal leptons  close or even inside hadronic jets. The details of mini-isolation can be found in Ref.~\cite{Rehermann:2010vq}, however for the purposes of the following paragraphs we briefly describe it as follows. In contrast to the standard lepton isolation requirement based on a fixed isolation cone radius $R^{\thinspace\text{cone}}\equiv \sqrt{\Delta \eta^2 + \Delta \phi^2}$, the mini-isolation criterium adopted in this work is based on defining a variable isolation cone radius
\begin{equation}
R^{\thinspace\text{cone}}\thinspace (p_{T\ell}) = \text{min}\Big(R_0,\ \frac{p_{T_0}}{p_{T_\ell}}\Big) \ ,
\label{mini1}
\end{equation}
where $R_0$ and $p_{T_0}$ are fixed parameters and $p_{T_\ell}$ is the transverse momentum of the candidate lepton. 

In order to illustrate the impact of lepton isolation on the four-top multi-lepton channels, we show in Fig.~\ref{leptons} the lepton multiplicity distributions before and after imposing isolation requirements in absence of kinematic cuts (apart from lepton isolation). Parton shower and hadronization effects are included via {\tt Pythia8}~\cite{Sjostrand:2014zea} and leptonic isolation cuts plus detector smearing are implemented with {\tt Delphes3}~\cite{deFavereau:2013fsa}. As we can see in the figure, the fraction of events with more than one lepton in the final state at truth-level (grey histogram) is significantly reduced once the standard isolation requirement is imposed (dashed red line). As stated above, many signal leptons fail to pass the standard isolation because at least one of the top-quarks is typically boosted {or because it accidentally ends up close to an unrelated jet}. We have also plotted in Fig.~\ref{leptons} the performance of the mini-isolation requirement (solid blue line) for the parameters in Eq.~\eqref{mini1} set to $R_0=0.5$ and $p_{T_0}=7.5$ GeV. This shows that the signal efficiencies for both the SS dilepton and trilepton channels can improve by a factor of $\sim2$ when compared to the standard isolation criteria. 

\begin{figure}[h!]
\begin{center}
\includegraphics[width=0.7\textwidth]{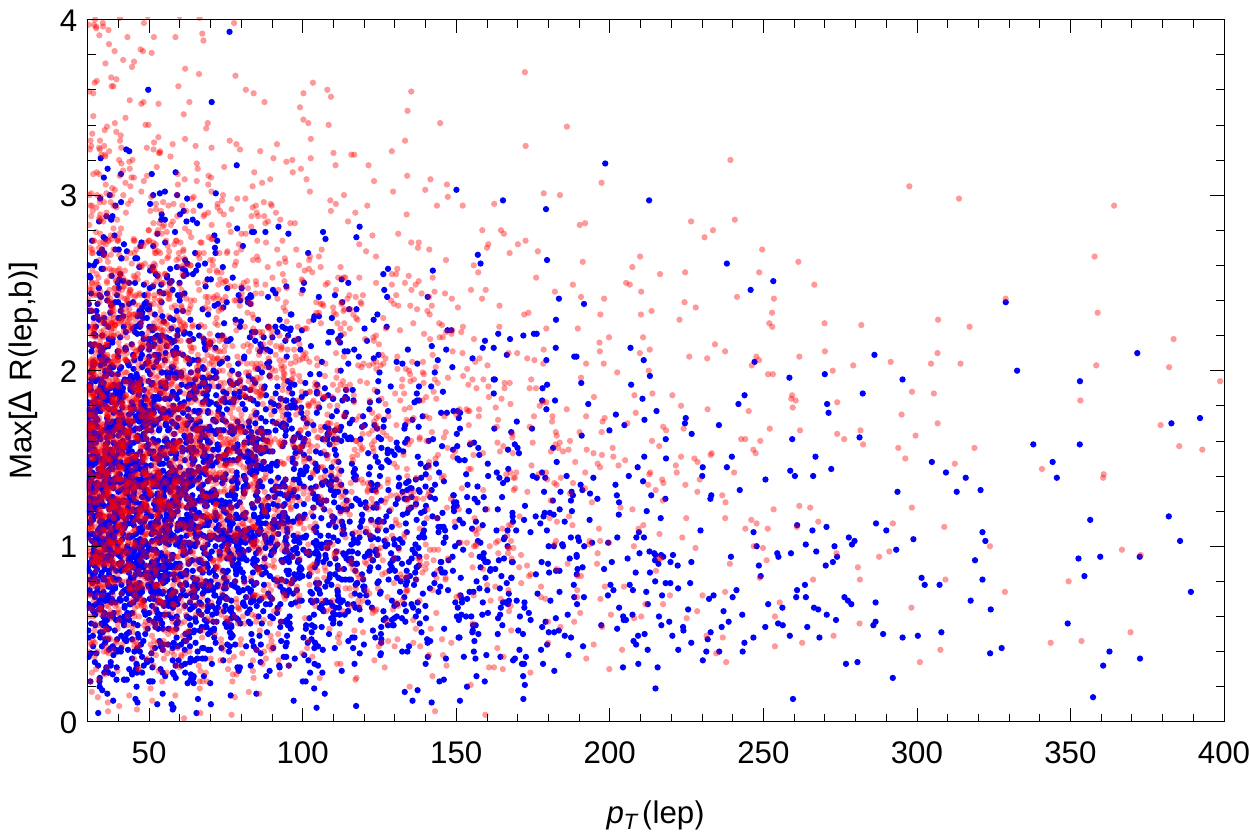}
\caption{
\small Distribution of signal (blue) versus background ($ttW^\pm b$ + jets, red) events for the maximum distance between a lepton and a $b$-quark as a function of the transverse momentum of the lepton.
}
\label{b-lep}
\end{center}
\end{figure}

As a last signal feature related to the presence of boosted top-quarks, we consider the  truth-level distance between leptons and $b$-quarks $\Delta R_{\ell b}$ as a function of the top-quark boost. As explained above, the lepton coming from a boosted top-quark decay is expected to be close to the $b$-quark. This is not the case for an energetic lepton arising from the leptonic decay of a $W$ boson in, e.g. the $t\bar tW^\pm$ background. For this reason functions of $\Delta R_{\ell b}$ could potentially work as good discriminants in a multi-lepton four-top search. In order to test this idea, we plot at parton level in Fig.~\ref{b-lep} the maximum separation $\Delta R^{\text{max}}_{\ell b}$ between each lepton and the set of $b$-quarks against the lepton transverse momentum for a sample of $pp\to\tttt$ decaying in the SS dilepton channel (blue dots) and compare it to a $t\bar tW^\pm$ background sample (red dots). Notice the mild separation between signal and background in the high-$p_T$ region, where signal events tend to cover the low $\Delta R^{\text{max}}_{\ell b}$ region while background events tend to cover larger $\Delta R^{\text{max}}_{\ell b}$ values. Unfortunately, the signal and background separation is not good enough and the expected number of events at LHC is too low, leading to a few-percent improvement of the signal significance for optimal choices of the cut on this variable. Nevertheless, at higher luminosities boost-sensitive angular discriminants such as $\Delta R_{\ell b}$ could eventually be exploited to improve sensitivity.

\section{Standard model four-top search strategy}
\label{section:3}

{The experimental program targeting the four-top signal at the LHC is currently in a preliminary stage. Both CMS and ATLAS collaborations have released a hand-full of dedicated searches for the SM four-top signal at the LHC. These are mainly based on the hadronic and mono-leptonic channels~\cite{Khachatryan:2014sca, Aad:2015kqa, ATLAS:2016gqb, CMS:2016wig}. Despite having much larger branching ratios than the multi-lepton decay modes, searches in these channels suffer from a tiny signal-to-background ratio due to overwhelming QCD and $t\bar t$ backgrounds. Better exclusion limits on the $\tttt$ cross-section have been obtained in a set of SUSY searches based on the SS dilepton signature~\cite{Zhou:2012dz, ATLAS:2016sno, Aad:2016tuk, CMS:2016vfu}. To our knowledge, the best upper bound to this date has been recently presented in a 13 TeV CMS search~\cite{CMS:2016vfu} and reads $\sigma_{\tttt}^{\text{SM}}< 57$\,fb. The signal regions constructed for these SUSY searches aim for NP signatures and as a consequence make use of strong cuts on kinematic variables such as the effective mass $m_{\text{eff}}$ and missing transverse energy $\slashed{E}_T$. The same is true for the recently proposed (resonant) $\tttt$ search strategy in the SS dilepton channel~\cite{Liu:2015hxi} which otherwise shares several features with our proposal. Unfortunately, these cuts are not optimal for the SM four-top signature at lower luminosities where signal events are far from abundant. Another problem with existing searches is that the SS dilepton and trilepton channels are usually combined during event selection in the same signal regions (e.g. see Ref.~\cite{ATLAS:2016sno}) and the differentiating features of the two signals with respect to their dominant backgrounds cannot be fully exploited.

In the following subsections we describe our SM four-top search strategy in the multi-lepton channels. In order to maximize signal sensitivity, we consider separately the SS dilepton and trilepton channels and show that each is optimized in a different signal region. Interestingly, for integrated luminosities ranging approximately between $50-75$ fb$^{-1}$, the trilepton channel surpasses in sensitivity the SS dilepton channel. The two searches described below avoid hard cuts on kinematic variables given that the four-top signal is statistically limited and also given the lack of a clear signal-to-background separation in the differential distributions (evidenced in, e.g. Fig.\ref{b-lep}). For this reason, both searches essentially consist in an optimization of jet and $b$-jet multiplicity selections. The main goal of this section is to demonstrate that by combining both dedicated multi-lepton searches, the LHC should achieve sensitivity to the SM four-top signal earlier than expected and possibly claim evidence of $\tttt$ production during Run-II. }

\subsection{Same-sign dilepton channel}
\label{section:3.1}

At hadron colliders, events with SS dileptons are rarely produced in the SM. On the other hand, a SS dilepton signal is predicted in a variety of beyond the SM scenarios such as universal extra-dimension models, supersymmetry and left-right symmetric models, making this signature a promising place to discover new physics at the LHC. Among the rare SM processes giving rise to SS dileptons, $pp\to\tttt$ is one of the sub-dominant contributions. Consequently, searches for SM $\tttt$ production in the SS dilepton channel need to discriminate the signal from much larger backgrounds. We performed a detailed classification of all irreducible and reducible backgrounds in Appendix \ref{samesignbackgrounds}. These fall into the following main categories: $t\bar tW$, $t\bar t Z$, $t\bar t H$ and "Others" for the irreducible backgrounds; in addition jets faking leptons and lepton charge mis-measurement ($Q$-flip) constitute the dominant reducible backgrounds.

We generated Monte-Carlo samples for both the signal and backgrounds using {\tt MadGraph} interfaced with {\tt Pythia8} to account for hadronization and showering effects. Background samples with large  jet multiplicities were generated with {\tt AlpGen}~\cite{Mangano:2002ea}, see Appendix \ref{samesignbackgrounds}. The detector response was simulated with {\tt Delphes3} in order to parametrically reproduce the LHC ATLAS detector. 

Our search strategy then proceeds as follows. For $b$-jet reconstruction, we emulate a high operating-point $b$-tagging algorithm with a reconstruction efficiency around $75\%$ for $b$-jets and rejection rates for charm and light jets of 8 and 400 respectively, see Ref.~\cite{ATL-PHYS-PUB-2015-022}. We isolate leptons using the mini-isolation requirement in Eq.~\eqref{mini1} with cone parameters $R_0=0.2$ and $p_{T_0}=8$ GeV for electrons, and $R_0=0.3$ and $p_{T_0}=10$ GeV for muons. In order to deal with isolated non-prompt leptons arising from heavy meson decays inside jets, we reject any isolated leptons within a distance of $\Delta R_{\ell j}<0.4$ from a reconstructed jet if the following condition is satisfied:
\begin{equation} 
p_{T_\ell} < \alpha\thinspace(\Delta R_{\ell j})^{\thinspace\beta}\ p_{T_{\text{jet}}}\,,
\label{nonprompt}
\end{equation}
\noindent where $p_{T_\ell}$ and $p_{T_{\text{jet}}}$ are the transverse momenta of the lepton and the jet, respectively.  The parameters in Eq.~\eqref{nonprompt} are fixed at $\alpha\!=\!0.19$ for electrons, $\alpha\!=\!0.17$ for muons and $\beta\!=\!-1$ for the angular exponent. A recent search by CMS~\cite{CMS:2016vfu}  used a similar condition to reduce isolated non-prompt backgrounds. For more details see Appendix~\ref{btag-NPL}. In addition, the following selection criteria are imposed on the reconstructed objects: only electrons  with $p_T>15$ GeV and $|\eta|<2$\footnote{This specific cut in $\eta$ removes large $e^\pm$ $Q$-flip background near the end-caps.}, muons with $p_T>10$ GeV and $|\eta|<2.5$, and jets of any flavor with $p_T>25$ GeV and $|\eta|<2.5$ are retained.

\begin{table}[t!]
\begin{center}
\begin{tabular}{l l l l l l }
\ $\mathcal{L}$=300 fb$^{-1}$ & \ \ & \bf\ SR6j\ & \bf\ SR7j\ & \bf\ SR8j\  \\
\hline
\\
\vspace{2pt}
$N_{\text{exp}}$ 	            & \ \ &139 (171) & 85 (101) & 43 (51)\\
\hline
$\mathbf{t\bar t t \bar t}$      & \ \ & \bf 16.7       &\bf13.5       &\bf 8.9 \\
$t \bar t W$            		   & \ \ & 60.7            & 35.0	   & 17.1 \\
$t \bar t Z$            		   & \ \ & 32.1            & 20.3          & 10.7 \\
$t \bar t h$            		   & \ \ & 5.5              & 3.1	           & 1.3 \\
Fakes                   		   & \ \ & 12.5 (17.3)  & 7.1 (9.8)	   & 3.3 (4.6)\\
Q-flip                      		   & \ \ &  7.6\  (34.4) & 3.7 (16.6)  & 1.6 (7.4)\\
Other                                    & \ \ &  4.4              & 2.4            & 1.0 \\
\hline
$\mathbf{S /B}$                    & \ \ &\bf 0.14  (0.11)  &\bf0.19  (0.15)& \bf 0.26 (0.21)\\
$\mathbf{S/\sqrt{B}}$            & \ \ &\bf 1.51  (1.34) &\bf1.60 (1.44)& \bf 1.53 (1.37)   \\
\hline
\end{tabular}
\caption{\small Four-top SS dilepton channel event yields assuming 300 fb$^{-1}$ at 13TeV LHC. Results are calculated using the benchmark $\epsilon_{\text{fake}}\!=\!7.2\times10^{-5}$ and $\epsilon_{\text{Qflip}}\!=\!2.2\times10^{-4}$, estimated in Appendix \ref{fakes}. In the parenthesis results are calculated using the conservative benchmark $\epsilon_{\rm fake}\!=\!10^{-4}$ and $\epsilon_{\text{Qflip}}\!=\!10^{-3}$. Here \textbf{SRnj} denotes the signal region with at least n tagged jets. The expected number of events is given by $N_{\text{exp}}=\text{Round}(S+B)$.}
\end{center}
\label{ss-results}
\end{table}

Events are then required to contain:
\begin{itemize}
\item Exactly one SS dilepton (events with additional leptons are vetoed).
\item Jet multiplicity (of any flavor) satisfying $N_j\ge6$.
\item $b$-jet multiplicity satisfying $N_b\ge3$.
\end{itemize}
Finally, we bin the selected events into signal regions, denoted by \textbf{SR{\em n}\! j}, and defined by different threshold values $n=6,7,8,...$ of the jet multiplicity $N_j \geq n$.

The results of such an analysis are shown in Table \ref{ss-results}. The expected number of total events $N_{\text{exp}}$ in each signal region as well as the expected number of events for the signal and each background category correspond to an integrated LHC luminosity of $\mathcal{L}=300$ fb$^{-1}$. {The reducible backgrounds have been estimated using two benchmark values for the fake lepton ($j\to\ell^\pm$) and $Q$-flip ($e^\mp\to e^\pm$) probabilities. The first benchmark defined by $\epsilon_{\text{fake}}\!=\!7.2\times10^{-5}$ and $\epsilon_{\text{Qflip}}\!=\!2.2\times10^{-4}$, is obtained by fitting MC simulations to existing 13 TeV results based on data-driven methods. All results are then calculated using these mis-identification probabilities. In addition, for illustration purposes, results inside the parenthesis are calculated using a set of more conservative benchmark values, $\epsilon_{\rm fake}\!=\!10^{-4}$ and $\epsilon_{\text{Qflip}}\!=\!10^{-3}$. For a detailed discussion on both benchmark choices see Appendix \ref{fakes}. }

For the SS dilepton channel the dominant background is the irreducible $t\bar t W$ background followed by $t\bar t Z$ and the $Q$-flip background. As expected, the signal-to-background ratio increases with the jet multiplicity and the signal sensitivity is maximized for {\bf SR7j} but drops for {\bf SR8j} due to limited statistics. Consequently, we find that using the projected LHC luminosity of $300\,{\rm fb}^{-1}$, a four-top search in the SS dilepton channel is expected to yield a signal significance around $S/\sqrt B=1.60$ with $S/B=0.19$ if at least 7 jets are selected in the final state. We note that dedicated experimental analyses are expected to outperform our chosen fake lepton and $Q$-flip probability benchmarks. In the limit, where the associated backgrounds can be completely neglected, the projected signal significance at $300\,{\rm fb}^{-1}$ luminosity improves to $S/\sqrt B=1.73$ with $S/B=0.22$ in the most sensitive {\bf SR7j} region.

\subsection{Trilepton channel}
\label{section:3.2}

The trilepton signature has been used in the past at the Tevatron~\cite{Abazov:2013eha,Aaltonen:2013vca} and more recently at the LHC~\cite{ATLAS-CONF-2016-003,CMS:2016dui} to search for SUSY and other beyond SM scenarios that predict much larger rates than the SM. Because it is more difficult to isolate three leptons, the four-top trilepton channel suffers from lower acceptance and efficiency when compared to the SS dilepton channel. This can be seen when comparing the bins $N_\ell=3$ and $N_\ell=2$ (SS) in Fig.~\ref{leptons}, where the ratio of events between the trilepton and the SS dilepton channels is predicted to be approximately $1:2$. This implies that the four-top trilepton channel may achieve a comparable sensitivity to the SS dilepton channel if their backgrounds in a given signal region satisfy a hierarchy of at least $1:4$.

The backgrounds for the trilepton channel have been classified in Appendix \ref{trilepbackgrounds}. In contrast to the SS dilepton signature, trileptons have the advantage of a lower instrumental background. The reason for this is (i) $Q$-flip is no longer a background, (ii) the SM processes producing fake trileptons have a lower rate when compared to those processes producing fake SS dileptons. This last point is evident if one keeps in mind that fake trileptons arise from mis-reconstructed $j\to\ell^\pm$ in $\ell^\pm\ell^\mp j$ while fake SS dileptons arise from mis-reconstruction of the much more abundant $\ell^\pm j$ final states. 

We apply the same lepton mini-isolation and $b$-tagging algorithm as in the SS dilepton case, as well as the condition in Eq.~\eqref{nonprompt} for rejecting non-prompt lepton backgrounds. We also use the same kinematic cuts for physical objects and select events according to the following criteria:

\begin{itemize}
\item Exactly three charged leptons (events with additional leptons are vetoed).
\item Jet multiplicity (of any flavor) satisfying $N_j\ge 4$.
\item b-jet multiplicity satisfying $N_b\ge 3$.
\item A $Z$-mass window veto: the invariant mass $m_{\ell\ell}$ of all possible same-flavor OS dileptons $\ell^+\ell^-$ must fall outside the mass window 70 GeV $ <m_{\ell\ell}<$ 105 GeV.
\end{itemize}

\noindent Selected events are then binned into signal regions \textbf{SR{\em n}\! j} with $n=4,5,6,...$,  defined in the same way as in the previous section.

\begin{table}[t!]
\begin{center}
\begin{tabular}{l l l l l l }
\ $\mathcal{L}$=300 fb$^{-1}$ & \ \ & \bf SR4j\ & \bf SR5j\ & \bf SR6j\  \\
\hline
\\
\vspace{2pt}
$N_{\text{exp}}$  	& \ \ &  31 (32)    & 25 (26)    & 17  (17)  \\
\hline
$\mathbf{t\bar t t \bar t}$            & \ \ & \bf 8.6      & \bf 7.8        & \bf 6.0 \\
$t \bar t Z$            			& \ \ &  9.9           & 8.0            &5.1  \\
$t \bar t W$            			 & \ \ &  6.7          & 4.9            &2.9  \\
$t \bar t h$            			 & \ \ &  2.3          & 1.8            & 1.2\\
Fakes                   			 & \ \ & 2.5 (3.5)   &  1.7 (2.4)  &0.9 (1.3)  \\
Other                                          & \ \ & 1.4            & 1.0           &  0.5\\
\hline
$\mathbf{S /B}$                           & \ \ &\bf 0.38 (0.36) &\bf 0.45 (0.43) & \bf  0.57 (0.54)\\
$\mathbf{S/\sqrt{B}}$                  & \ \ &\bf 1.80 (1.76) &\bf 1.87 (1.84) & \bf  1.84 (1.80) \\
\hline 
\end{tabular}
\caption{\small  Four-top trilepton channel event yields assuming 300 fb$^{-1}$ at 13TeV LHC. Results are calculated using the benchmark value $\epsilon_{\text{fake}}\!=\!7.2\times10^{-5}$ estimated in Appendix \ref{fakes}. In the parenthesis results are calculated using the conservative benchmark value $\epsilon_{\rm fake}\!=\!10^{-4}$.}
\end{center}
\label{trilep-results}
\end{table}

We present the results for the trilepton analysis in Table \ref{trilep-results} for a projected LHC luminosity of 300 fb$^{-1}$. The $Z$-mass veto used in this analysis has a signal acceptance of approximately 90\% and a very large rejection rate of the otherwise overwhelming $t\bar t Z$ background. This cut is thus instrumental in making the trilepton channel competitive with the SS dilepton channel despite smaller signal event rates. For the signal region \textbf{SR5j} we obtain a maximum significance of $S/\sqrt{B}=1.87$ with a high signal-to-background ratio of $S/B=0.45$. In case the jet to lepton fake-rate can be significantly reduced compared to our conservative benchmark value $\epsilon_{\rm fake}=10^{-4}$ (see Appendix~\ref{fakes}), the significance can be improved up to $S/\sqrt{B}=1.97$ with a high signal-to-background ratio of $S/B=0.50$.

\subsection{Results}
\label{section:3.3}

We are now in position to give results for the SM four-top search in the combined SS dilepton and trilepton channels. We implemented a statistical analysis based on a log-likelihood test \cite{Cowan:2010js} using the most sensitive signal regions for each channel, i.e. \textbf{SR7j} for the SS dilepton channel and \textbf{SR5j} for the trilepton channel. Our analysis is based on the following main assumptions: (i) For the total number of observed events we use the sum of the expected signal and total background events. (ii) We only consider uncertainties for the dominant backgrounds, i.e. $\sim12$\% and $\sim13$\% of theoretical uncertainties (on the NLO cross-sections) for $t\bar t Z$ and $t\bar t W$ respectively \cite{Alwall:2014hca} and a $\sim50$\% uncertainty fixed at $\mathcal{L}_{\text{int}}=13.2$ fb$^{-1}$ for the Fakes and $Q$-flip backgrounds extracted from Ref.~\cite{ATLAS:2016kjm}. For each uncertainty we assign an independent nuisance parameter following a gaussian prior. (iii) In order to project our results to arbitrary luminosities, the number of signal and background events are scaled with $\mathcal{L}_{\text{int}}$, while the statistical uncertainties for the Fakes and $Q$-flip backgrounds are scaled with $\sqrt{\mathcal{L}_{\text{int}}}$. (iv) Estimated backgrounds for the Fake leptons and $Q$-flip are based on the mis-identification probabilities $\epsilon_{\text{fake}}\!=\!7.2\times10^{-5}$ and $\epsilon_{\text{Qflip}}\!=\!2.2\times10^{-4}$ estimated in Appendix~\ref{fakes}.

\begin{figure}[t!]
\begin{center}
\includegraphics[width=0.45\textwidth]{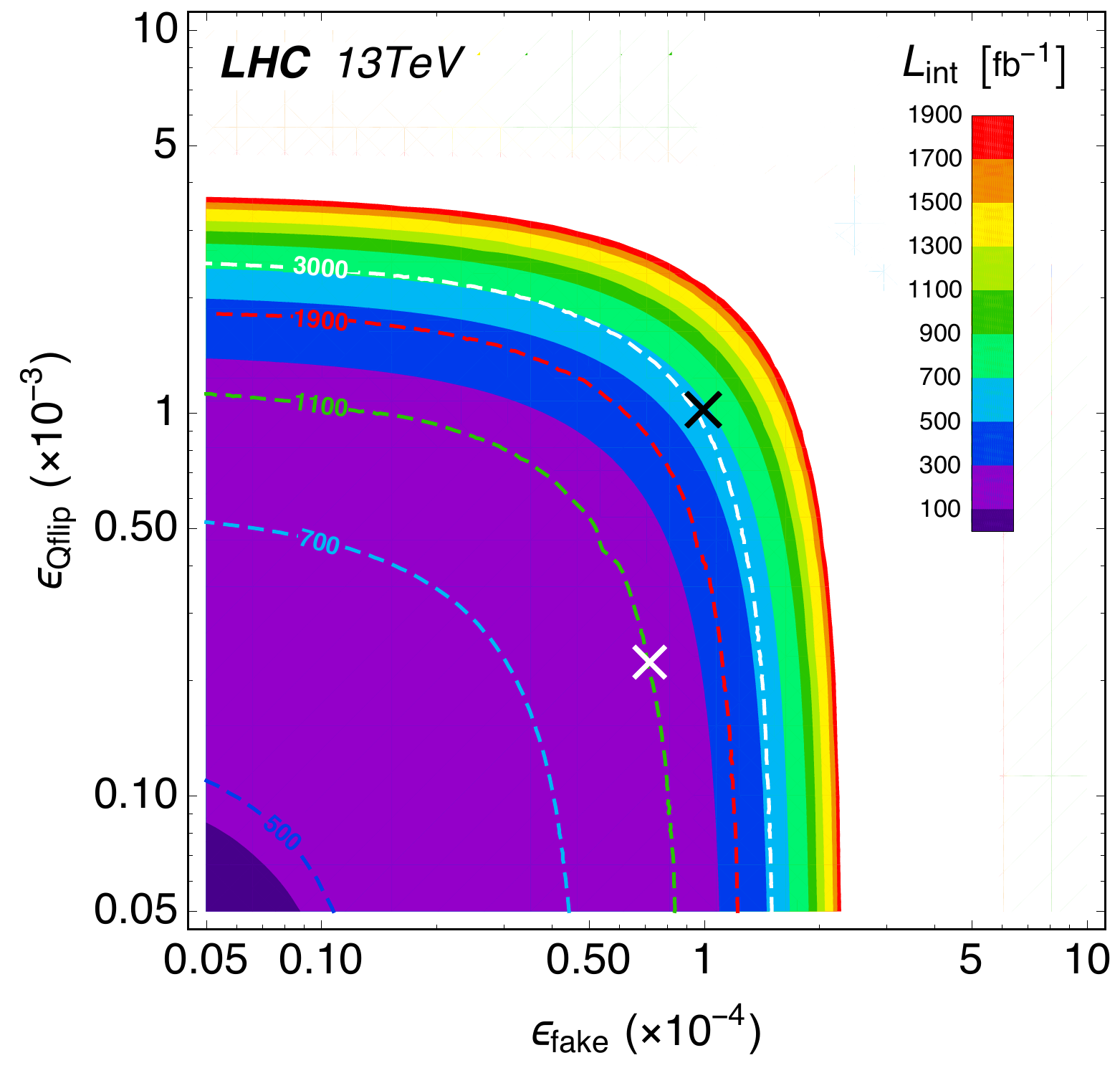}~~
\includegraphics[width=0.428\textwidth]{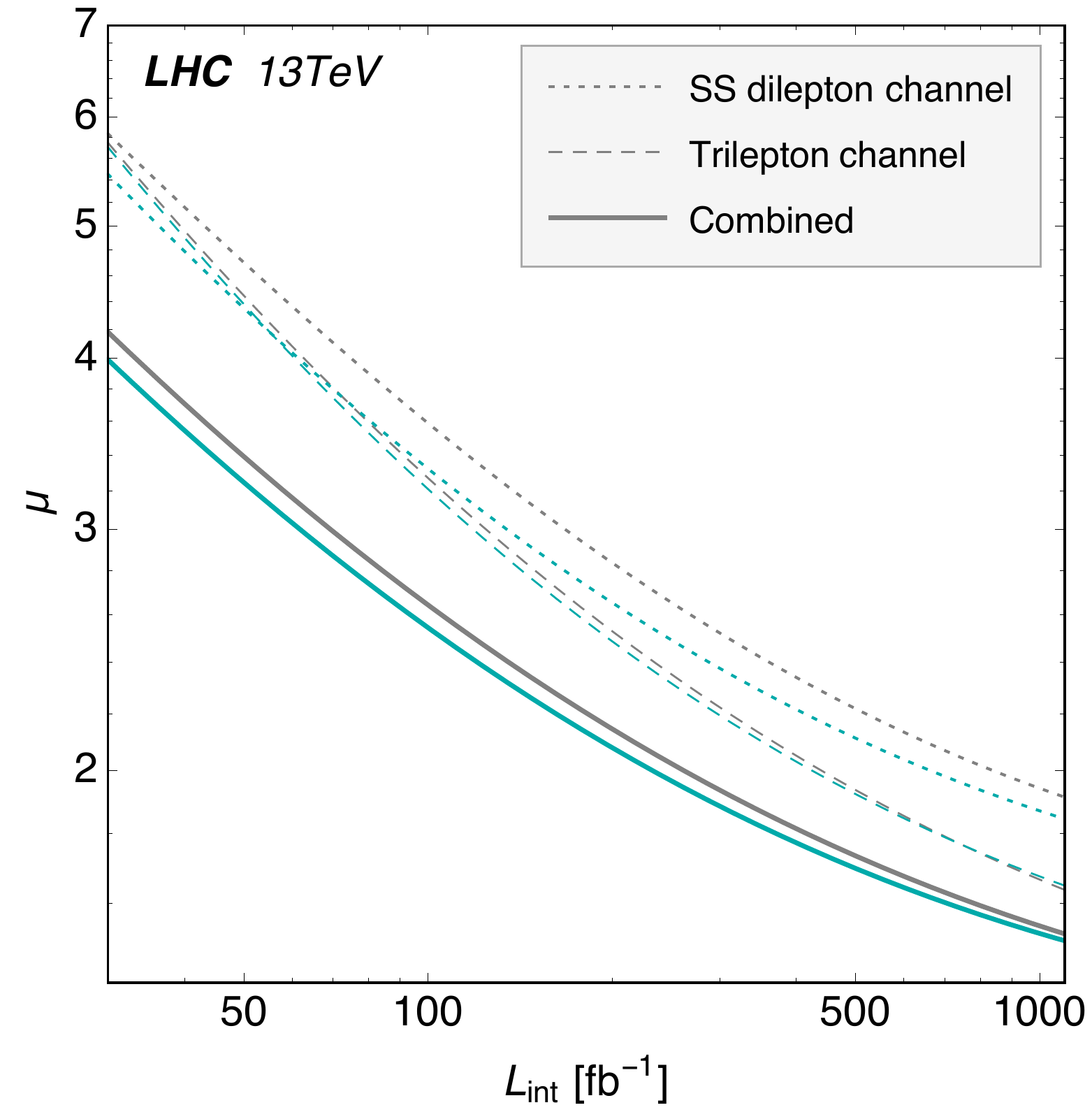}
\caption{\small  (Left panel) Iso-contours of $3\sigma$ evidence luminosities (fb$^{-1}$)  (shaded contours) and $5\sigma$ discovery luminosities (fb$^{-1}$) (dashed lines) in the $(\epsilon_{\text{fake}}, \epsilon_{\text{Qflip}})$-plane of lepton-faking and $Q$-flip probabilities. The white cross corresponds to the estimated values $(7.2\times10^{-5},2.2\times10^{-4})$ from Appendix~\ref{fakes}, while the black cross corresponds to the conservative benchmark values $(10^{-4},10^{-3})$. (Right panel) Projected ATLAS 95\% CL exclusion limits for the SM four-top signal strength as a function of the integrated luminosity for the SS dilepton channel (blue dotted line), the trilepton channel (blue dashed line) and the combination (blue solid line). We also include the results obtained from reducible backgrounds estimated using the conservative benchmark (gray lines).
\small
}
\label{results}
\end{center}
\end{figure}

We now present {\it evidence} and {\it discovery} luminosities for the SM four-top signal based on our search strategy, i.e. predict when the LHC should exclude the background-only hypothesis at $3\sigma$ and $5\sigma$ respectively. We find that the LHC experiments should be able to establish evidence ($p$-value of $3\times10^{-3}$) for the SM four-top process at a combined integrated luminosity of $\mathcal{L}_{\text{int}}\simeq215$ fb$^{-1}$, and claim discovery ($p$-value of $3\times10^{-7}$) at a higher integrated luminosity of $\mathcal{L}_{\text{int}}\simeq1060$ fb$^{-1}$. These results show that with our search strategy, the LHC starts becoming sensitive to four-top production in the SM at luminosities achievable within Run-II.

As already alluded to above, these projections are very sensitive to the Fake and $Q$-flip background estimations, which are based on particular set of values for the lepton-faking and $Q$-flipping probabilities. To illustrate this point, we  plot in Fig.~\ref{results} (Left) iso-contours of both $3\sigma$ evidence (shaded contours) and  $5\sigma$ discovery (dashed lines) luminosities in the $(\epsilon_{\text{fake}}, \epsilon_{\text{Qflip}})$-plane. The mis-identification probabilities used for our results is marked with a white cross, while the much more conservative benchmark $\epsilon_{\text{fake}}\!=\!10^{-4}$ and $\epsilon_{\text{Qflip}}\!=\!10^{-3}$ is marked with a black cross. For example, improving upon our estimates for $\epsilon_{\rm fake,Qflip}$ by an order of magnitude would make the associated backgrounds completely negligible. In that limit we find that the LHC experiments should be able to establish $3\sigma$ evidence for SM $\tttt$ production already at an integrated luminosity of $\mathcal{L}_{\text{int}}\simeq98$ fb$^{-1}$, and claim discovery at $\mathcal{L}_{\text{int}}\simeq420$ fb$^{-1}$. 

The extracted upper limit from our search for the SM four-top signal strength is
\begin{equation} \label{uperbound}
\mu^{\text{SM}}_{\tttt}\ \le\ 1.87\ \ \ \text{at 95\% CL, }
\end{equation}
for an integrated luminosity of $\mathcal{L}_{\text{int}}=300$ fb$^{-1}$. In Fig.~\ref{results} (Right) we give the 95\% CL exclusion limits on the signal strength $\mu^{\text{SM}}_{\tttt}$ for each individual multi-lepton channel (blue dotted and dashed lines) and for both combined (blue solid line) as a function of the integrated luminosity. We also present results (gray lines) obtained using the benchmark  $\epsilon_{\text{fake}}\!=\!10^{-4}$ and $\epsilon_{\text{Qflip}}\!=\!10^{-3}$ to estimate the reducible backgrounds. Notice that the extracted upper bound for the signal strength in the combined analysis does not change considerably when using this more conservative benchmark. As expected, for higher luminosities the trilepton channel achieves better sensitivity than the SS dilepton channel and drives the search.


\section{Application to new physics}
\label{section:5}

Many NP models addressing the SM hierarchy or flavor puzzles predict new TeV scale dynamics coupling most strongly to the third generation, in particular the top quark (see e.g. Refs.~\cite{Pomarol:2008bh, Kane:2011zd}). Such interactions can most naturally be searched for through multi-top-quark production. Several existing proposals target pair- or $t\bar t$-associated production of heavy resonances decaying to top quark pairs~\cite{Gerbush:2007fe,Acharya:2009gb,Cacciapaglia:2011kz,Perelstein:2011ez, AguilarSaavedra:2011ck,Gregoire:2011ka, Liu:2015hxi, Gori:2016zto}. In this case the dominant signature is the appearance of resonances in $t\bar t$ invariant mass spectra. In all cases, for multi-TeV resonance masses, boosted top searches can be effective in these scenarios~\cite{Kim:2016plm}.

On the other hand, color-neutral particles coupling predominantly to the third generation with masses below the $t\bar t$ threshold are at present only weakly constrained~\cite{Haisch:2016hzu}. Such states appear in models addressing recent $B$-meson decay anomalies~\cite{Fajfer:2012jt, Faroughy:2016osc} or in scenarios of cosmological thermal relic dark matter (see e.g. Ref.~\cite{Dolan:2014ska, Arina:2016cqj}). The exchange of such particles mediating four-top production would generically result in kinematics, not strikingly different from the dominant QCD contributions. Their dominant effect is thus expected to be a modification of the inclusive four-top production cross-section . We study this possibility in more detail using two representative toy model examples in which we extend the SM with respectively a new vector or scalar boson affecting the four-top production at the tree-level: (1) a top-philic neutral $Z'$ vector boson and (2) a neutral scalar $\phi$ with Yukawa couplings to the top. In the first model we assume for simplicity that the $Z'$ with mass $m_{Z'}$ only couples significantly to right-handed top quarks\footnote{This same interaction was studied in Ref.~\cite{Liu:2015hxi}, but in a different parameter region.}. The relevant interaction Lagrangian then reads
\begin{equation}
\mathcal L_{Z'} = - g_{tZ'} \bar t_R \slashed Z' t_R\,.
\end{equation}
We note that the chiral top-current, to which $Z'$ is coupled is broken explicitly by the top quark mass (and by anomalies) and the $m_{Z'}\to 0$ limit cannot be approached trivially in this model. Nonetheless, well defined UV completions exist in the literature where these issues are properly addressed with no immediate consequences for $\tttt$ phenomenology (see e.g. Refs.~\cite{Jackson:2009kg,Jackson:2013rqp,Goertz:2015nkp} for an explicit example as well as Ref.~\cite{Kamenik:2011vy} for a more general discussion).   
In the second model, the relevant $\phi-t$ interactions are on the other hand described by
\begin{equation}
\mathcal L_{\phi} = - y_{t\phi} \bar t_L \phi t_R + \rm h.c.\,.
\label{eq:phiModel}
\end{equation}
Depending on the phase of $y_{t\phi}$, this interaction is in general CP violating. While the form of interactions above is not manifestly invariant under the SM EW gauge symmetry, suitable UV completions in terms of multiple Higgs doublet or singlet SM extensions can be easily constructed where the dominant effects in $\tttt$ phenomenology are captured by the effective Lagrangian in Eq.~\eqref{eq:phiModel} (see e.g. Ref.~\cite{Dev:2014yca}). By choosing $m_\phi=m_h=125$~GeV, this second example also covers the interesting case of a modified top Yukawa coupling of the SM Higgs boson. 
\begin{figure}[t!]
\begin{center}
\includegraphics[width=0.45\textwidth]{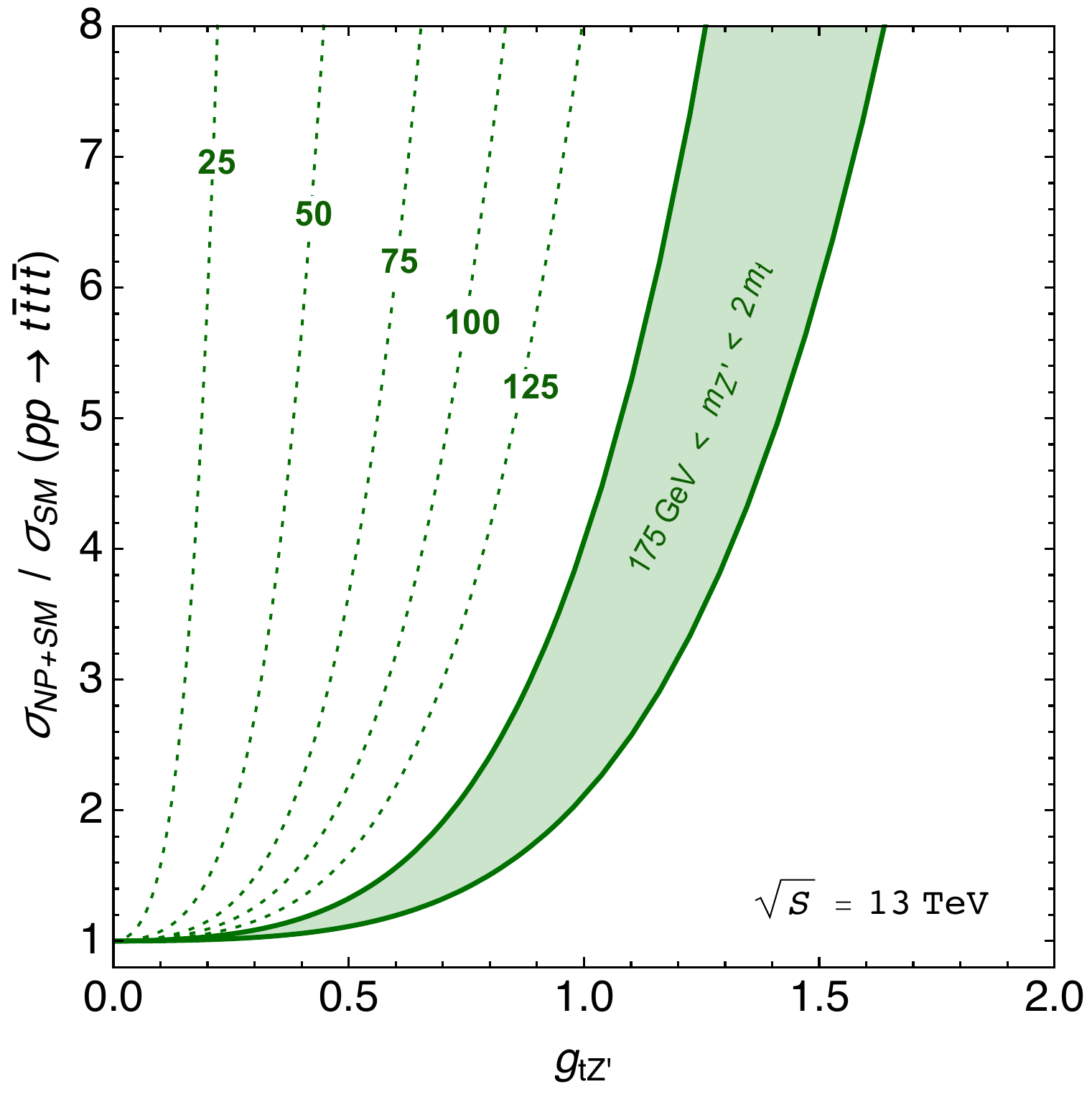}
\includegraphics[width=0.45\textwidth]{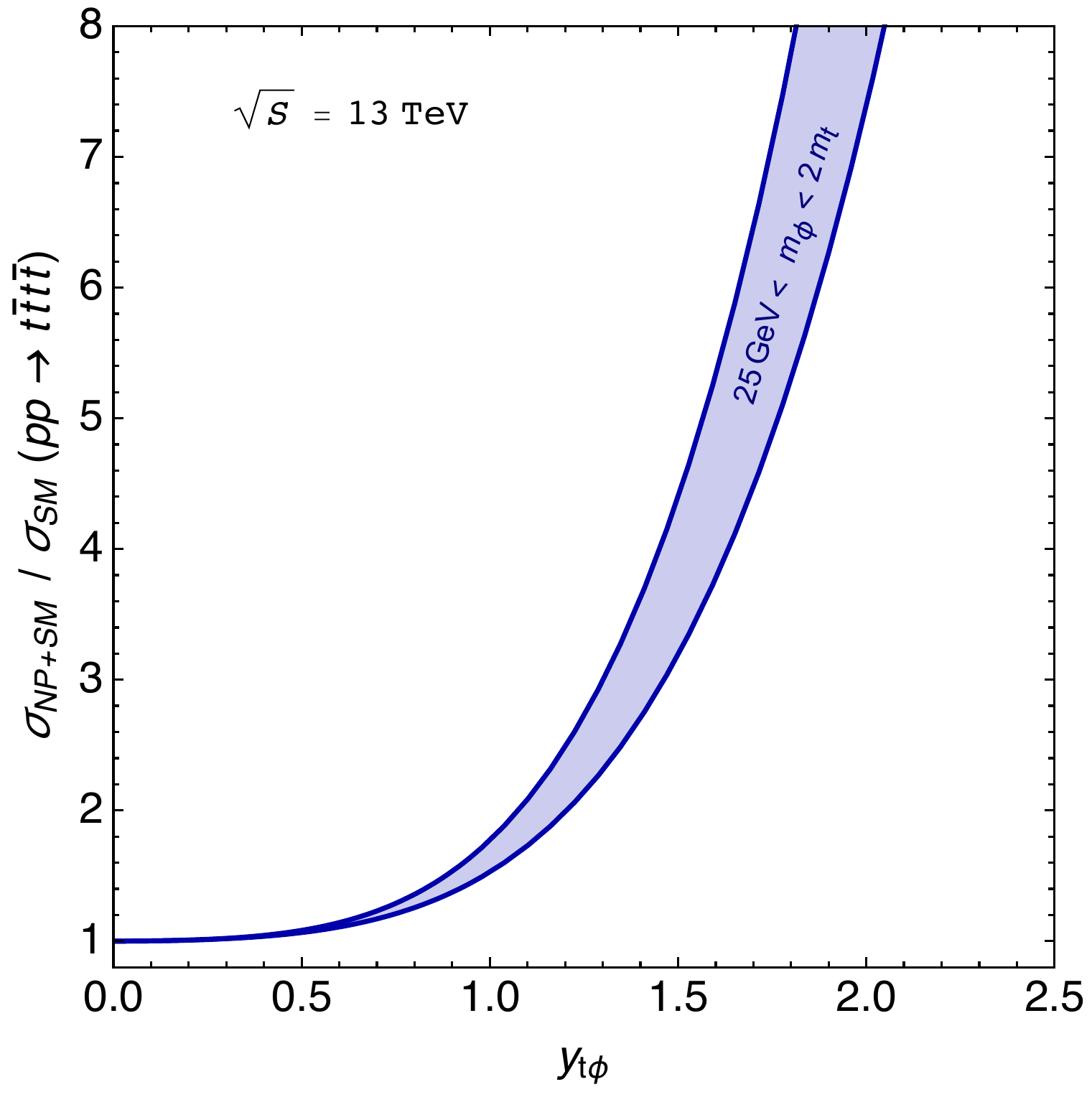}
\caption{\small
\small Predictions for the deviation $\sigma_{NP+SM}/\sigma_{SM}$ in the $pp\to \tttt$ cross-section at $\sqrt{s}=13$ TeV within the simplified NP $Z^{\prime}$ (left-hand side) and $\phi$ (right-hand side) models as a function of the couplings $g_{tZ'}$ and $y_{t\phi}$, for different $Z'$ and $\phi$ masses, respectively.}
\label{NP}
\end{center}
\end{figure}

In both models, since the new degrees of freedom are never produced on-shell, their effects in four-top production are largely independent of their possible other interactions. \footnote{This should be compared to Ref.~\cite{ Liu:2015hxi}, where the new particle is assumed to be heavier than $2m_t$, appearing as an on-shell resonance that decays to $t \bar t$, and consequently motivating and resulting in a somewhat different search strategy.} In our case, contributions can be parametrized in terms of the mediator mass and the relevant coupling to top quarks, ($m_{Z'}, g_{tZ'}$) for model (1) and ($m_{\phi}$,$y_{t\phi}$) for model (2), respectively, and in particular do not depend on the mediator decay width. This is in contrast to direct $t\bar t$ resonance searches, where the resonance width can play an important role (see e.g.~\cite{Dicus:1994bm, Gori:2016zto, Carena:2016npr, Craig:2016ygr}).  

\begin{figure}[t!]
\begin{center}
\includegraphics[width=0.45\textwidth]{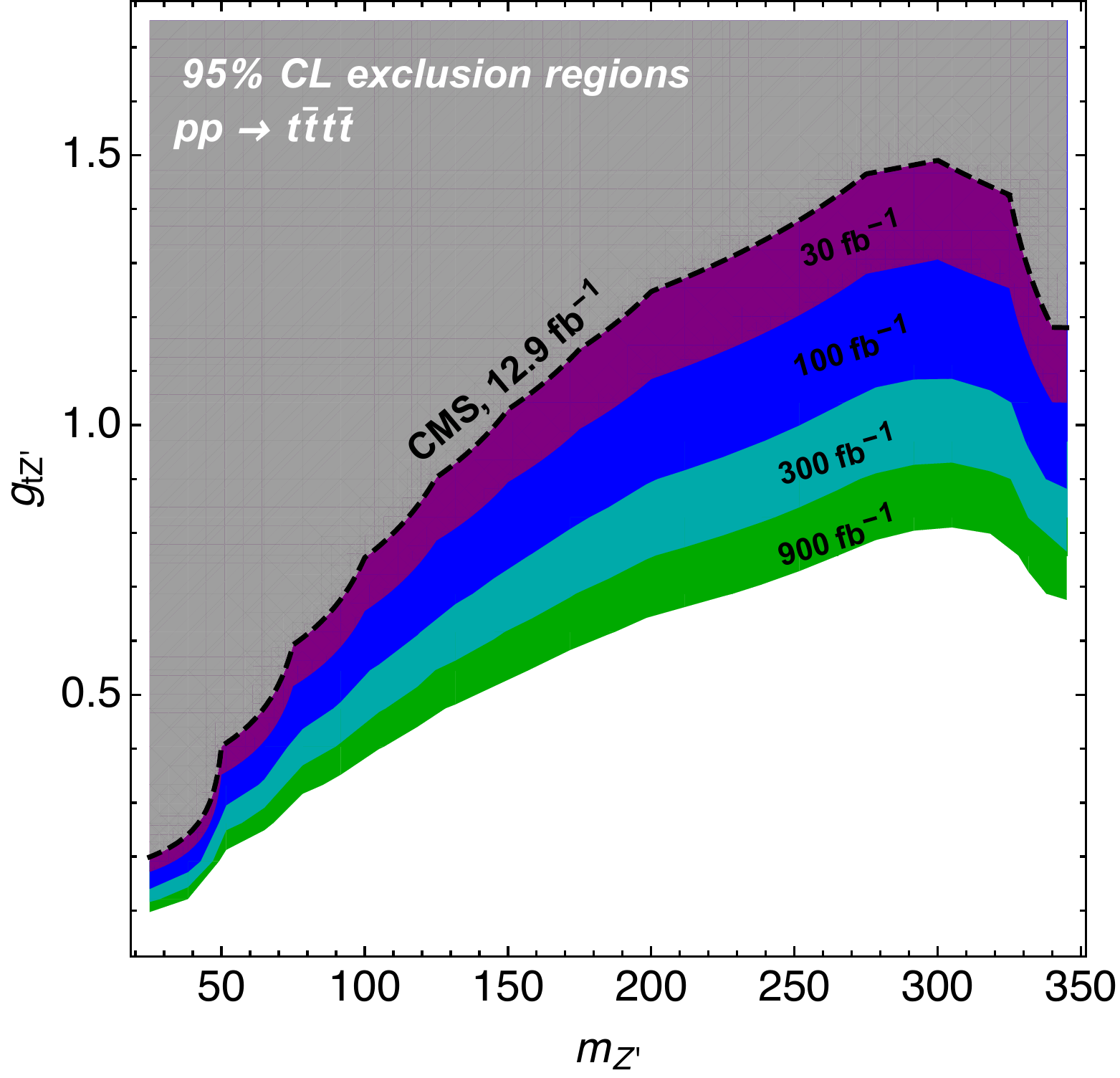}~
\includegraphics[width=0.45\textwidth]{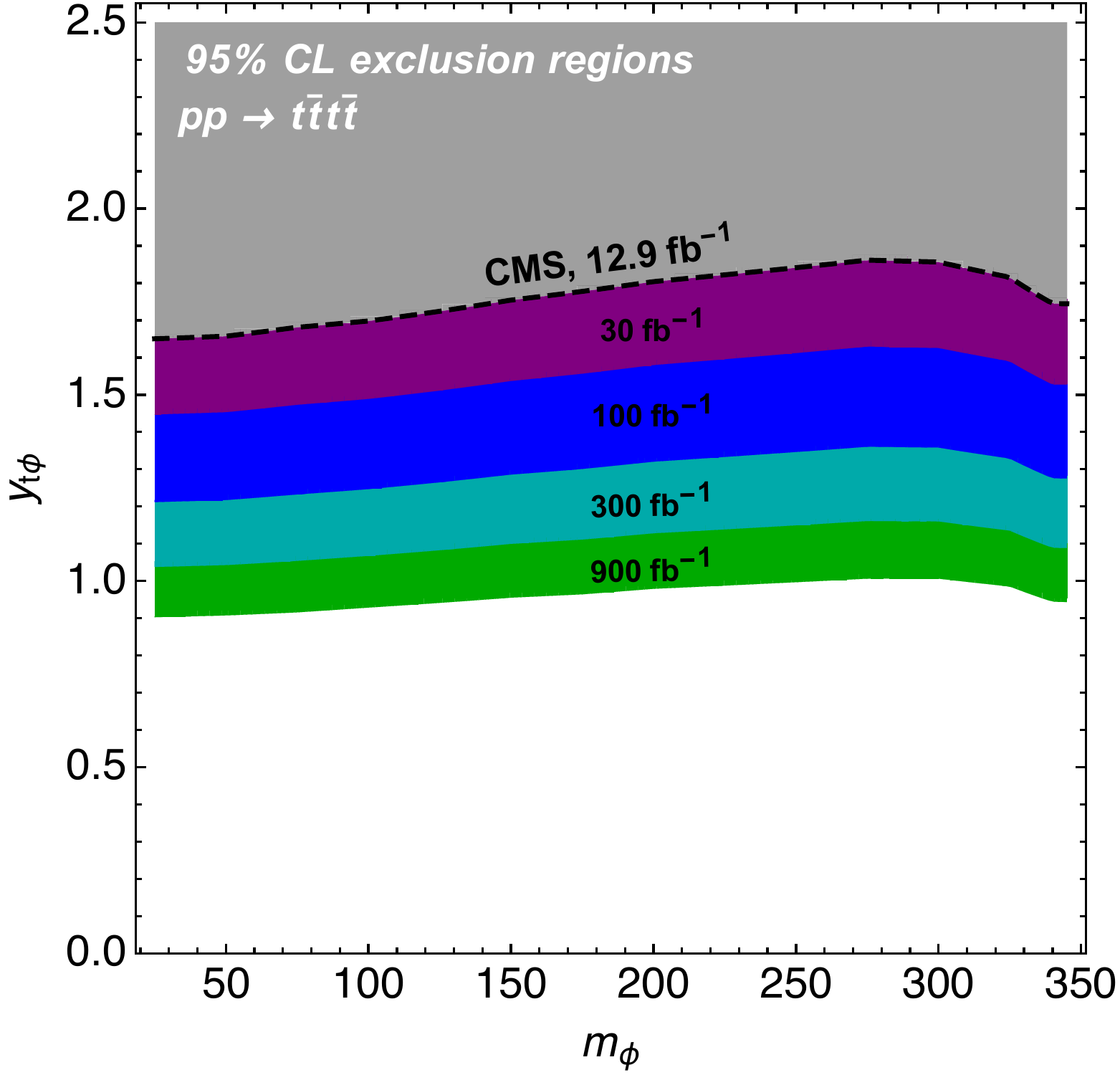}
\caption{\small NP exclusion regions for LHC luminosities of 30fb$^{-1}$ (Purple), 100 fb$^{-1}$ (Blue), 300 fb$^{-1}$ (Dark Cyan) and 900 fb$^{-1}$ (Green) respectively, for the $Z^{\prime}$ model (Left) and the $\phi$ scalar model (Right). The existing bound extracted from the recent CMS search~\cite{CMS:2016vfu} is shown in gray shade and bounded by a dashed contour.
}
\label{NPexclusion}
\end{center}
\end{figure}

In Fig.~\ref{NP} we show the predicted four-top production cross-section including NP contributions $\sigma_{NP+SM}(pp\to\tttt)$ in both models, normalized to the SM cross-section prediction $\sigma_{SM}(pp\to\tttt)$, all computed at LO in QCD. We find that for both NP models, the off-shell mediator contributions to $pp\to\tttt$ can considerably enhance the four-top production at the LHC. For the top-philic $Z^\prime$ model (left panel), the enhancement with respect to the SM cross-section becomes almost independent of the ${Z'}$ mass in the range $m_t\lesssim m_{Z^\prime}\lesssim 2m_t$ being roughly a factor of two for couplings satisfying $0.5\lesssim g_{tZ^\prime}\lesssim 1$. On the other hand, the strong $Z'$ mass dependence for $m_{Z'}\ll m_t$ can be easily understood since the $Z'$ couples to an unconserved current and the scattering amplitudes receive contributions proportional to its breaking due to $m_t$, and thus grow with $m_t/m_{Z'}$.  For the neutral scalar model (right panel) the enhancement in the cross-section is practically mass-independent in the whole considered $\phi$ mass range. Here, one can expect an enhancement of order $\sigma_{NP+SM}/\sigma_{SM}\sim 2$ for couplings of order $y_{t\phi}\sim1$. In both models, for mediator masses close to $t\bar t$ threshold, interference effects make the mass dependence non-monotonous. We note that the sign of these interference effects is fixed irrespective of the sign (or phase) of $g_{tZ'} (y_{t\phi})$, since the NP mediated amplitudes are proportional to $g^2_{tZ'} (|y_{t\phi}|^2)$ in both models, respectively.  Finally, resonance width effects (not included in our analysis) are expected to become relevant in the region a few GeV below threshold. 

In the mass region of interest, the main effect of the considered NP contributions is to rescale the four-top cross-section while leaving the main kinematic distributions SM-like. Indeed, we have verified that the acceptances and efficiencies of the NP contributions for both proposed multi-lepton four-top search strategies are approximately constant in all of the NP parameter space and comparable to the SM signal. We are therefore able to apply the upper limits on the SM four-top production extracted from our search in the previous section directly. The resulting 95\% CL exclusion regions in the NP parameter space (mass vs. coupling) for both simplified models are shown in Fig.~\ref{NPexclusion}. For comparison, we also show the corresponding bounds extracted from a recent CMS search~\cite{CMS:2016vfu}. Finally we note that, while we do not project our constraints to larger mediator masses, where the NP mediated $\tttt$ production is resonantly enhanced, we expect our search strategy to retain good sensitivity also in this region of the NP models' parameter space, as long as the resonant $t\bar t$ pairs are not significantly boosted ($m_{Z',\phi} \slashed{\gg} 2 m_t$).


\section{Conclusions and outlook}
\label{section:6}

The LHC will soon start exploring a new realm of rare SM processes. In case no new signal is detected in the most common search channels at the LHC during Run-II, rare processes such as $pp\to\tttt$ will become the next generation of probes for physics beyond the SM. These SM signatures can easily receive sizeable contributions from  new dynamics otherwise transparent to existing experimental probes. For this reason it is important to have a set of simple yet effective dedicated search strategies aiming to measure these challenging signatures and probe possible deviations from SM predictions.

In this work we have presented a simple search strategy for detecting four top-quarks produced in the SM at the LHC. In order to avoid large backgrounds in the all-hadronic and mono-leptonic decay modes, we have focused on the much cleaner SS dilepton and trilepton channels. In both cases, the final state is highly populated with light-jets, $b$-jets and leptons with enough energy to produce considerable overlap and merging between some of the final states, resulting in a substantial drop in signal efficiency (specially due to non-isolated leptons). We have also shown that significant separation between the four-top signal and the main backgrounds (such as $t\bar t W^\pm$ and $t\bar t Z$) can be achieved using the $b$-jet and jet multiplicity distributions. These signal features have been exploited in order to maximize $S/B$ and $S/\sqrt{B}$ for each channel, resulting in signal categories based on the mini-isolation requirement for leptons and a large b-jet and jet multiplicity cuts ($N_j\ge7$ for the SS dilepton channel, $N_j\ge5$ for the trilepton channel, and $N_b\ge3$ for both). For the trilepton case, we have included a $Z$-mass veto (in the OS and same-flavor dilepton invariant mass) in order to control the dominant $t\bar t Z$ backgrounds and substantially increase signal-to-background ratio and signal significance. Interestingly, this allows for the trilepton channel to achieve a better four-top sensitivity during LHC Run-II than the SS dilepton channel. In the Appendices, we give a detailed classification of all necessary irreducible and reducible backgrounds relevant for our analysis. 

Projection results for evidence and discovery luminosities are given in Sec.~\ref{section:3.3}. We have found that by combing the SS dilepton and trilepton channels, SM four-top production should be evidenced ($3\sigma$) at 215 fb$^{-1}$ of data, and possibly discovered ($5\sigma$) around 1000 fb$^{-1}$. These results are based on current estimations of the reducible fake-lepton and $Q$-flip backgrounds, extracted from fitting both mis-identification probabilities to 13 TeV data-driven predictions by ATLAS. These mis-identification probabilities have been estimated to be $\epsilon_{\text{fake}}\!=\!7.2\times10^{-5}$ for fake lepton ($j\to\ell^{\pm}$) and $\epsilon_{\text{Qflip}}\!=\!2.2\times10^{-4}$ for $Q$-flip ($e^\pm\to e^\mp$), as explained in Appendix~\ref{fakes}.  We  give a projected upper limit on the $\tttt$ signal strength (cross-section normalized to the SM value) of about 1.87 at 95\% CL for 300 fb$^{-1}$ of data.  We have also analyzed the impact on the luminosities required for evidence, discovery and (95\% CL) upper limits for four-top production when different fake lepton and $Q$-flip rates are considered for reducible background estimation (see Fig.~\ref{results}). In the optimistic scenario where the experimental collaborations manage to substantially reduce fake leptons and $Q$-flip backgrounds to negligible levels, four-top production may be evidenced in the very near future, at approximately 100 fb$^{-1}$, and discovered at approximately 420 fb$^{-1}$. 

The SM four-top search strategy presented above can also be used to probe non-resonant NP models with SM-like kinematics affecting four-top production, as for instance, a color-neutral new particle mainly coupling to top-quarks and with a mass below the $2 m_t$ threshold. This scenario would indeed be invisible to most of the available four-top searches presented so far~\cite{ATLAS:2016sno,Aad:2016tuk,CMS:2016ahn,Khachatryan:2014sca,Zhou:2012dz}. We have used the constrains derived from our multi-lepton search for SM four-top production to set limits on two simplified top-philic models: a neutral scalar boson and a $Z^\prime$ model. We have shown that our search can cover important regions of parameter space in both models at current luminosities and can probe couplings bellow $\sim1$ with 900 fb$^{-1}$ of data.

As a final comment, we discuss some possible ways to improve our mutli-lepton search strategy. We have described in Sec.~\ref{section:2} some features that could, in principle, help distinguish the SM four-top signal from the dominant background source $t\bar t V$. For instance, a lower cut on the effective mass $m_{\text{eff}}$, being significantly larger for the four-top final state when compared to $t\bar t V$, could be used as a good signal-to-background discriminant at higher luminosities. Another interesting kinematic variable is the minimum distance between leptons and $b$-jets (shown at truth-level in Fig.~\ref{b-lep}) as well as possibly other variables based on angles between leptons and $b$-jets. A third possibility, which could be implemented in case of an energy upgrade, is to include harder rapidity cuts for leptons and $b$-jets by exploiting the fact that the $t\bar t V$ backgrounds will tend to be produced more into the forward region than the more massive $\tttt$ signal (already at 13 TeV, we estimated a 10\% increase in signal significance at truth-level). Relevant especially in the high luminosity regime (beyond Run-II and Run-III), we leave it to the experimental collaborations to include (if necessary) these additional kinematical cuts on top of the multi-lepton four-top search strategy described in this work. Finally, the analysis presented here has been carried out by applying the same operating point (OP) in the rejection-efficiency plane for all the $b$-tags in the event selection. However, the signal significance might be improved with a multiple OP selection. We have found that using three different OPs among the ones presented in Ref.~\cite{ATL-PHYS-PUB-2015-022}, it is possible to achieve up to one order of magnitude enhancement on the effective rejection for a given efficiency compared to the case where the same OP is applied to all the b-tags in the event. Likewise, multiple OPs may allow to improve the effective efficiency up to $10\ \%$ for a given rejection.


\section*{Acknowledgments}

E.A.~and J.F.K.~thank participants in Voyages Beyond the SM workshop for useful discussions. A.S., E.A., and R.M.~work was partially supported by ANPCyT PICT 2013-2266. J.F.K. acknowledges the financial support from the Slovenian Research Agency (research core funding No. P1-0035).


\vspace{10pt}

\begin{table}[h!]
\begin{center}
\begin{tabular}{|l |l |c |r |l |r |c |}
\hline
Category & Backgrounds & FS & $\sigma$ [fb] & decay mode & $\sigma \times BR$ [fb] & comments \\
\hline
$t\bar t W$ & $t \bar{t}\ W^{\pm} $                             & 5 &   350.4   & $W_{\ell^\pm}\ W_{\ell^\pm}\ W_{\text{had}}$     & 16.84    &       \\
                  & $t \bar{t}\ W^{\pm}\ j$              	& 5 &  167.8    & $W_{\ell^\pm}\ W_{\ell^\pm}\ W_{\text{had}}$	   & 8.06     & MLM   \\
                  & $t \bar{t}\ W^{\pm}\ jj$              	& 5 &  96.8     & $W_{\ell^\pm}\ W_{\ell^\pm}\ W_{\text{had}}$     & 4.65     & MLM   \\
                  & $t \bar{t}\ W^{\pm}\ jj$              	& 5 &  \ \           & $W_{\ell^\pm}\ W_{\ell^\pm}\ W_{\ell^\mp} $ & 1.58     & MLM, lost $\ell$  \\
                  & $t \bar{t}\ W^{\pm}\  bjj$            	& 5 & 2.3          & $W_{\ell^\pm}\ W_{\ell^\pm}\ W_{\text{had}}$  & 0.11     &       \\
                  & $t \bar{t}\ W^{\pm}\ b\bar b\ jj$     	& 4 &  2.1         & $W_{\ell^\pm}\ W_{\ell^\pm}\ W_{\text{had}}$          & 0.10     &       \\
\hline
$t\bar t Z$  & $t \bar{t}\ Z $                     & 5  &   583.3         & $W_{\ell^\pm}\ W_{\text{had}}\ Z_{\ell}$     & 22.33   & lost $\ell$  \\
                  & $t \bar{t}\ Z\ j$              & 5  &   404.7         & $W_{\ell^\pm}\ W_{\text{had}}\ Z_{\ell}$     & 15.50   & MLM, lost $\ell$   \\
                  & $t \bar{t}\ Z\ jj$             & 5  &    194.9        & $W_{\ell^\pm}\ W_{\text{had}}\ Z_{\ell}$     & 7.46    & MLM, lost $\ell$ \\
                  & $t \bar{t}\ Z\ jj$             & 5  &  \ \            & $W_{\ell^\pm}\ W_{\ell^\pm}\ Z_{\ell}$       & 3.18    & MLM, lost $\ell$   \\
\hline
$t\bar t h$ & $t \bar{t}\ h $                      & 4  &  397.6       & $W_{\ell^\pm}\ W_{\text{had}}\ W_{\ell^\pm}\ W_{\text{had}}$       &  4.70 &  $h\to WW^*$  \\
                 & $t \bar{t}\ h $                 & 4  &         \ \  & $W_{\ell^\pm}\ W_{\text{had}}\ Z_{\ell}\ Z_{\text{had}}$           & 0.37   &  $h\to ZZ^*$  \\
                 & $t \bar{t}\ h $                 & 5  &  401.3       & $W_{\ell^\pm}\ W_{\text{had}}\ \tau_{\ell^\pm}\ \tau_{\text{had}}$ & 2.18   &  $h\to \tau^+\tau^-$   \\
\hline
Others           & $t Z\ bjj$                      & 5  & 176.7       & $W_{\ell^\pm}\ Z_{\ell}$                                        & 4.52    & lost $\ell$ \\
                 & $t \bar{t}\ W^+W^-$             & 4  &  8.0        & $W_{\ell^\pm}\ W_{\text{had}}\ W_{\ell^\pm}\ W_{\text{had}}$    & 0.57    &    \\
                 & $t \bar{t}\ W^+W^-$             & 4  &  \ \        & $W_{\ell^\pm}\ W_{\text{had}}\ W_{\ell^+}\ W_{\ell^-}$          & 0.39    &  lost $\ell$  \\
                 & $W^\pm W^\pm\ b \bar b j j$     & 4  &  1.25       & $W_{\ell^\pm}\ W_{\ell^\pm}$                                    & 1.94    &    \\
                 & $ZZ\ b\bar{b} j$                & 4  &  30.2       & $Z_{\ell}\ Z_{\ell}$                                            & 0.31    &  lost $\ell$  \\
\hline
\bf Signal      & $\tttt$                          & 4 &  \bf 9.2   & $W_{\ell^\pm}\ W_{\ell^\pm}\ W_{\text{had}}\ W_{\text{had}}$      & \bf 0.66    &    \\
\hline
\end{tabular}
\caption{\small Irreducible backgrounds for the SS dilepton search. In the comment column, "MLM" indicates that the jet matching was performed. "lost $\ell$" implies that for this background to produce a SS dilepton one or more of the leptons in a multi-lepton final state is lost either by not satisfying isolation requirements or down the beam pipe. In the last row we have included for comparison the SM four-top signal in the SS dilepton decay mode.}
\label{SSback}
\end{center}
\end{table}

\begin{table}[t!]
\begin{center}
\begin{tabular}{|l |l |c |r |l |r | }
\hline
Category & Backgrounds & FS & $\sigma$ [pb] & decay mode & $\sigma \times BR\times\epsilon$ [fb]  \\
\hline
Fake          &$t \bar{t}\, j$              &5& 301.6 & $W_{\ell^\pm} W_{\text{had}} $ & 11.43  \\
              &$t \bar{t}\, jj$             &5& 124.9 & $W_{\ell^\pm} W_{\text{had}} $ & 4.74  \\
              &$t \bar{t}\, b jj$           &5& 5.3 & $W_{\ell^\pm} W_{\text{had}} $   & 0.20  \\
              &$t \bar{t}\,b \bar{b}\, jj$  &4& 3.0 & $W_{\ell^\pm} W_{\text{had}} $   & 0.11  \\
              &$t \bar{t}\,b \bar{b}\, 3j$  &4& 2.3 & $W_{\ell^\pm} W_{\text{had}}$    & 0.09  \\
\hline
$Q$-flip &$t \bar{t}\, jj$                   &5& 124.9 & $W_{\ell^\pm} W_{\ell^\mp} $ & 8.03 \\
              &$t \bar{t}\, b jj$            &5& 5.3   & $W_{\ell^\pm} W_{\ell^\mp} $ & 0.34  \\
              &$t \bar{t}\,b \bar{b}\, jj$   &4& 3.0   & $W_{\ell^\pm} W_{\ell^\mp} $ & 0.19  \\
              &$t \bar{t}\,b \bar{b}\, 3j$   &4& 2.3   & $W_{\ell^\pm} W_{\ell^\mp} $ & 0.15  \\
              &$Z\,b \bar{b}\, 2j$           &4& 26.3  & $Z_\ell$                     & 2.66  \\
\hline
\end{tabular}
\caption{\small Reducible backgrounds for the SS dilepton search. Here $\epsilon=10^{-4}$ for the Fake category and $\epsilon=10^{-3}$ for $Q$-flip category.}
\label{SSRedback}
\end{center}
\end{table}

\section*{Appendix}
\label{section:8}

\renewcommand{\thesubsection}{\Alph{subsection}}

\subsection{Breakdown of SS dilepton backgrounds}
\label{samesignbackgrounds}

In this appendix we describe the backgrounds used for the analysis of Sec.~\ref{section:3.1}. The backgrounds that mimic the four-top SS dilepton signature fall into two types: (i) {\it Irreducible backgrounds} coming from rare SM processes that contain one real SS dilepton pair in the final state, (ii) {\it Reducible backgrounds} of instrumental origin where a "fake" SS dilepton pair is produced in the detector. 

All background simulations were performed at LO with {\tt MadGraph} for samples with less than two extra partons in the final state or {\tt AlpGen} for samples with more than two extra partons. We chose the NN23LO set for the parton distribution function (PDF) and {\tt FastJet} \cite{Cacciari:2011ma,Cacciari:2005hq} for clustering jets using the anti-$k_T$ algorithm with $R=0.4$. Background samples with up to 2 extra partons at the matrix element (ME) level were merged and matched  to the parton shower using the MLM matching scheme. When available~\cite{Alwall:2014hca}, the LO cross-sections were rescaled with a $K$-factor to include NLO corrections. 

\paragraph{Irreducible backgrounds}
The non-negligible background processes used in the SS dilepton search strategy are divided into background categories as shown in the two first columns of Table~\ref{SSback}. There, we considered as an irreducible background any process producing final states at the ME level: one SS dilepton and a minimum of 4 hard quarks of which at least 2 must be $b$-quarks. In the two following columns we give the production cross-section calculated either in the 4-flavor or 5-flavor schemes (FS)\footnote{The definition of the jet $j$ in the second column depends on the FS: $j$ arises from the partons $g,u,d,s,c$ in the 4FS and from $g,u,d,s,c,b$ in the 5FS.}, subject to the partonic cuts $p^j_T>25$ GeV and $\Delta R_{jj}>0.2$ for jets of any flavor. All irreducible background processes eventually decay into an intermediate boson state $V=W,Z$ (either on or off the mass shell), denoted by $V_{\ell}$ or $V_{\text{had}}$ depending on whether these decay leptonically or hadronically into final states fermions. The decay modes into the final state fermions are shown in the fifth column, while on the sixth column we give a rough estimate of the total cross-section of the background process based on the corresponding branching ratios $BR$. In the last column we give additional comments. We have checked that other backgrounds, such as $tVV$ and $VVV$, etc ($V=W,Z,h$) are negligible when demanding $N_b\ge3$ and did not include them in the classification.

\paragraph{Reducible backgrounds}
Due to a large contribution from fake leptons and charged-flipped lepton pairs, any four-top analysis at the LHC in the multi-lepton channel must include a correct estimation of these background processes. In this work, we estimate the fake lepton and charge-flip ($Q$-flip) backgrounds by simulating all potential SM sources giving rise to one fake SS dilepton at the detector level via jet-to-lepton faking $j\to\ell^\pm$ or a $Q$-flip for electron/positrons only $e^\pm\to e^\mp$. These mis-identifications are mainly produced through the following mechanisms:
\begin{itemize}

\item Fake leptons: Can originate from incorrectly tagging as a lepton a charged meson related to a final state parton $j$. Another main source included in this category are non-prompt (real) leptons originating from the leptonic decay of a heavy meson inside a final state jet $j$. In this way an initial $\ell^\pm j$ pair can be detected as a SS dilepton via $j\to\ell^\pm$ mis-identification.

\item $Q$-flip: Can occur when a real OS dilepton pair coming from some underlying process is mistaken in the detector as a SS dilepton by mis-identifying the charge of one of the leptons. This effect is negligible for muons given that their charge is measured in both the inner silicon tracker and at the outer muon detector layer. Physically, the dominant contribution comes from trident conversion inside the detector: an incoming electron/positron suffers bremsstrahlung $e^\pm\to \gamma e^\pm\to e^\mp e^\pm e^\pm$ inside the silicon tracker. Since the probability of $Q$-flip is proportional to the flight length inside the medium, the central region of the tracker (e.g. $|\eta|<2$) produce less $Q-$flipped electrons than regions near the end-caps. 
\end{itemize}

The dominant sources generating both reducible backgrounds are $t\bar t$ + jets followed by $W^\pm$ + jets and $Z$ + jets. For simplicity, it is enough to assume that the probabilities $\epsilon_{\text{fake}}$ and  $\epsilon_{\text{Qflip}}$ for these mis-identifications are flat in $p_T$ and $\eta$ within the kinematic regions of our search strategy. We also assume that the fake lepton or $Q$-flipped electron (positron) inherit exactly the same kinematic properties as the source jet or positron (electron). Under these simplifications we just need to weight each of the simulated source samples with the corresponding mis-identification probabilities, here fixed at the benchmark values $\epsilon_{\text{fake}}=10^{-4}$ and  $\epsilon_{\text{Qflip}}=10^{-3}$ (see Appendix~\ref{fakes}). The complete list of all dominant  background sources generating the SS dilepton reducible backgrounds are given in Table \ref{SSRedback}.

\begin{table}[b!]
\begin{center}
\begin{tabular}{|l |l |c |r |l |r |c |}
\hline
Category & Backgrounds & FS & $\sigma$ [fb] & decay mode & $\sigma \times BR$ [fb] & comments \\
\hline
$t\bar t W$ & $t \bar{t}\ W^{\pm}\ jj$              	& 5 &  96.8   & $W_{\ell^\pm}\ W_{\ell^\pm}\ W_{\ell^\pm}$     & 1.58  & MLM  \\
                  & $t \bar{t}\ W^{\pm}\  bjj$         	& 5 &   2.3   & $W_{\ell^\pm}\ W_{\ell^\pm}\ W_{\ell^\pm}$     & 0.04   &            \\
                  & $t \bar{t}\ W^{\pm}\ b\bar b\ jj$ 	& 4 &   2.1  & $W_{\ell^\pm}\ W_{\ell^\pm}\ W_{\ell^\pm}$      & 0.03  &            \\
\hline
$t\bar t Z$  & $t \bar{t}\ Z $                          & 5  &   583.3  & $W_{\ell^\pm}\ W_{\text{had}}\ Z_{\ell}$     & 22.33  &     \\
                  & $t \bar{t}\ Z\ j$                   & 5  &   404.7  & $W_{\ell^\pm}\ W_{\text{had}}\ Z_{\ell}$     & 15.50  & MLM    \\
                  & $t \bar{t}\ Z\ jj$                  & 5  &    194.9  & $W_{\ell^\pm}\ W_{\text{had}}\ Z_{\ell}$    & 7.46  & MLM    \\
                  & $t \bar{t}\ Z\ jj$                  & 5  &      & $W_{\ell^\pm}\ W_{\ell^\pm}\ Z_{\ell}$           & 3.18   & MLM, lost $\ell$    \\
\hline
$t\bar t h$ & $t \bar{t}\ h $                           & 4  & 397.6  &$W_{\ell^\pm}\ W_{\ell^\pm}\ W_{\ell^\mp}\ W_{\text{had}}$           &1.60  &  $h\to WW^*$  \\
                 & $t \bar{t}\ h $                      & 4  &     & $W_{\ell^\pm}\ W_{\ell^\mp}\ Z_{\ell}\ Z_{\text{had}}$                 & 0.06 &  $h\to ZZ^*$  \\
                 & $t \bar{t}\ h $                      & 5  &   401.3  & $W_{\ell^\pm}\ W_{\ell^\mp}\ \tau_{\ell^\pm}\ \tau_{\text{had}}$  & 0.74 &  $h\to \tau^+\tau^-$   \\
\hline
Others      & $t\ Z\ bjj$                               & 5  &  176.7 & $W_{\ell^\pm}\ Z_{\ell}$                                            & 4.52    &  \\
                 & $W^\pm Z\ b\bar b\ jj $              & 4  & 70.3   & $W_{\ell^\pm}\ Z_{\ell}$                                            & 1.80    &  \\
                 & $t \bar{t}\ W^+W^-$                  & 4  & 8.0   & $W_{\ell^\pm}\ W_{\ell^\pm}\ W_{\ell^\mp}\ W_{\text{had}}$           & 0.39    &  \\
                 & $ZZ\ b\bar{b} j$                     & 4  & 30.2   & $Z_\ell\ Z_\ell$                                                    & 0.31    &  \\                                  
\hline
\bf Signal      & $\tttt$                               & 4 &  \bf9.2   & $W_{\ell^\pm}\ W_{\ell^\pm}\ W_{\ell^\mp}\ W_{\text{had}}$        & \bf 0.45  &    \\
\hline
\end{tabular}
\caption{\small Irreducible backgrounds for the trilepton search. In the comment column, "MLM" indicates that the jet matching was performed. "lost $\ell$" implies that for this background to produce a trilepton one lepton from a four-lepton final state is lost either by not satisfying isolation requirements or down the beam pipe. In the last row we have included for comparison the SM four-top signal in the trilepton decay mode.}
\label{3Lback}
\end{center}
\end{table}

\begin{table}[h!]
\begin{center}
\begin{tabular}{|l |l |c |r |l |r |}
\hline
Category & Backgrounds & FS & $\sigma$ [pb] & decay mode & $\sigma \times BR\times\epsilon$ [fb]  \\
\hline
Fake          &$t \bar{t}\, jj$               &5& 124.9 & $W_{\ell^\pm} W_{\ell^\mp} $ & 0.80 \\
              &$t \bar{t}\, b jj$             &5& 5.3 & $W_{\ell^\pm} W_{\ell^\mp} $   & 0.03  \\
              &$t \bar{t}\,b \bar{b}\, jj$    &4&  3.0 & $W_{\ell^\pm} W_{\ell^\mp} $  & 0.02  \\
              &$t \bar{t}\,b \bar{b}\, 3j$    &4&  2.3 & $W_{\ell^\pm} W_{\ell^\mp} $  & 0.01  \\
              &$Z\,b \bar{b}\, 2j$            &4& 26.3 & $Z_\ell$                      & 0.27  \\
\hline
\end{tabular}
\caption{\small Reducible backgrounds for the trilepton search. Here $\epsilon=10^{-4}$ for the Fake category.}
\label{3LRedback}
\end{center}
\end{table}

\subsection{Breakdown of trilepton backgrounds}
\label{trilepbackgrounds}

We now turn to the dominant backgrounds for the trilepton channel used in Sec.~\ref{section:3.2}. The classification is very similar to the SS dilepton and is given in Table \ref{3Lback} for the irreducible backgrounds and Table \ref{3LRedback} for the reducible backgrounds. All simulations were performed using the same tools and settings as described in Appendix~\ref{samesignbackgrounds}.


\subsection{Fake lepton and $Q$-flip mis-identification probabilities}
\label{fakes}

\begin{figure}[t!]
\begin{center}
\includegraphics[width=0.45\textwidth]{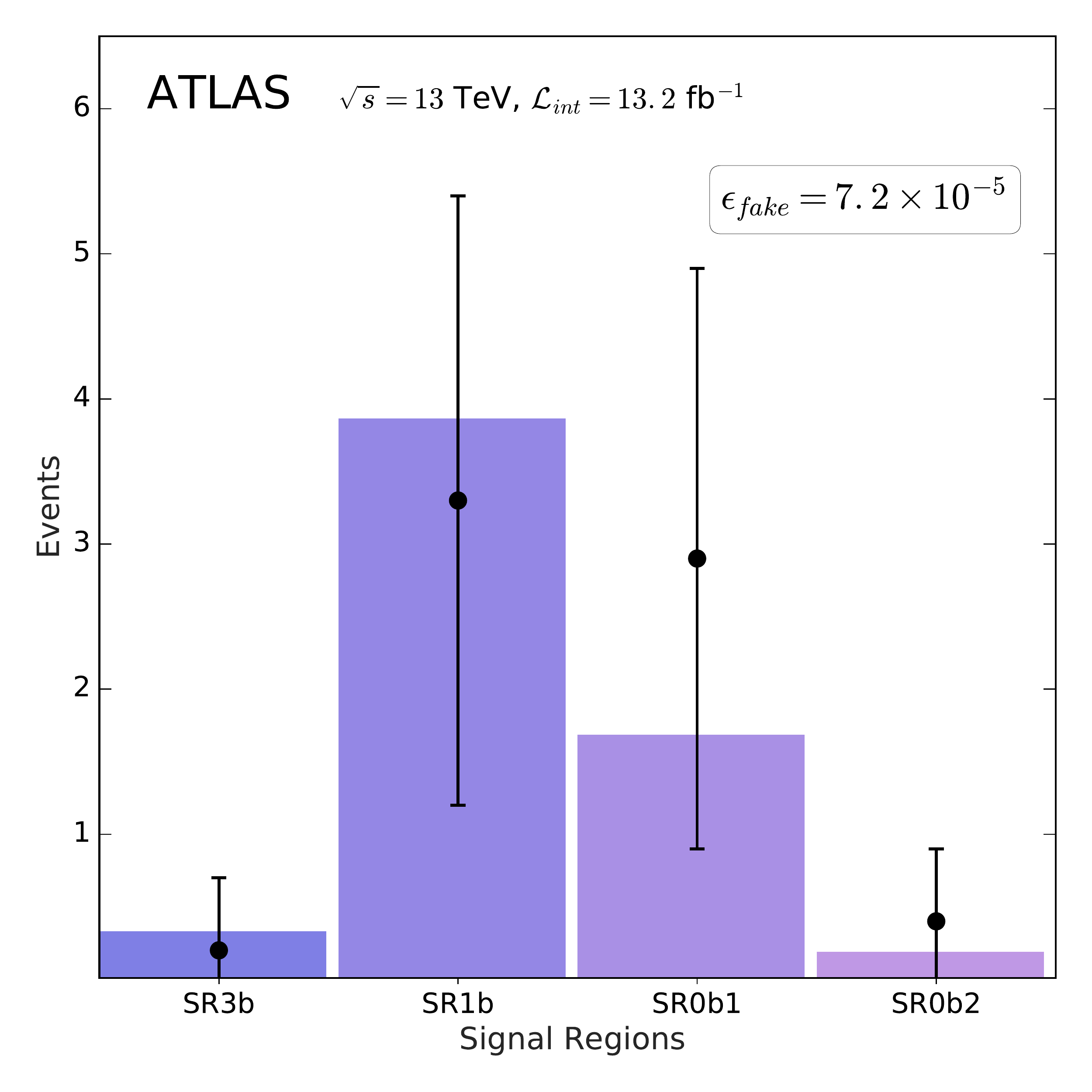}
\includegraphics[width=0.45\textwidth]{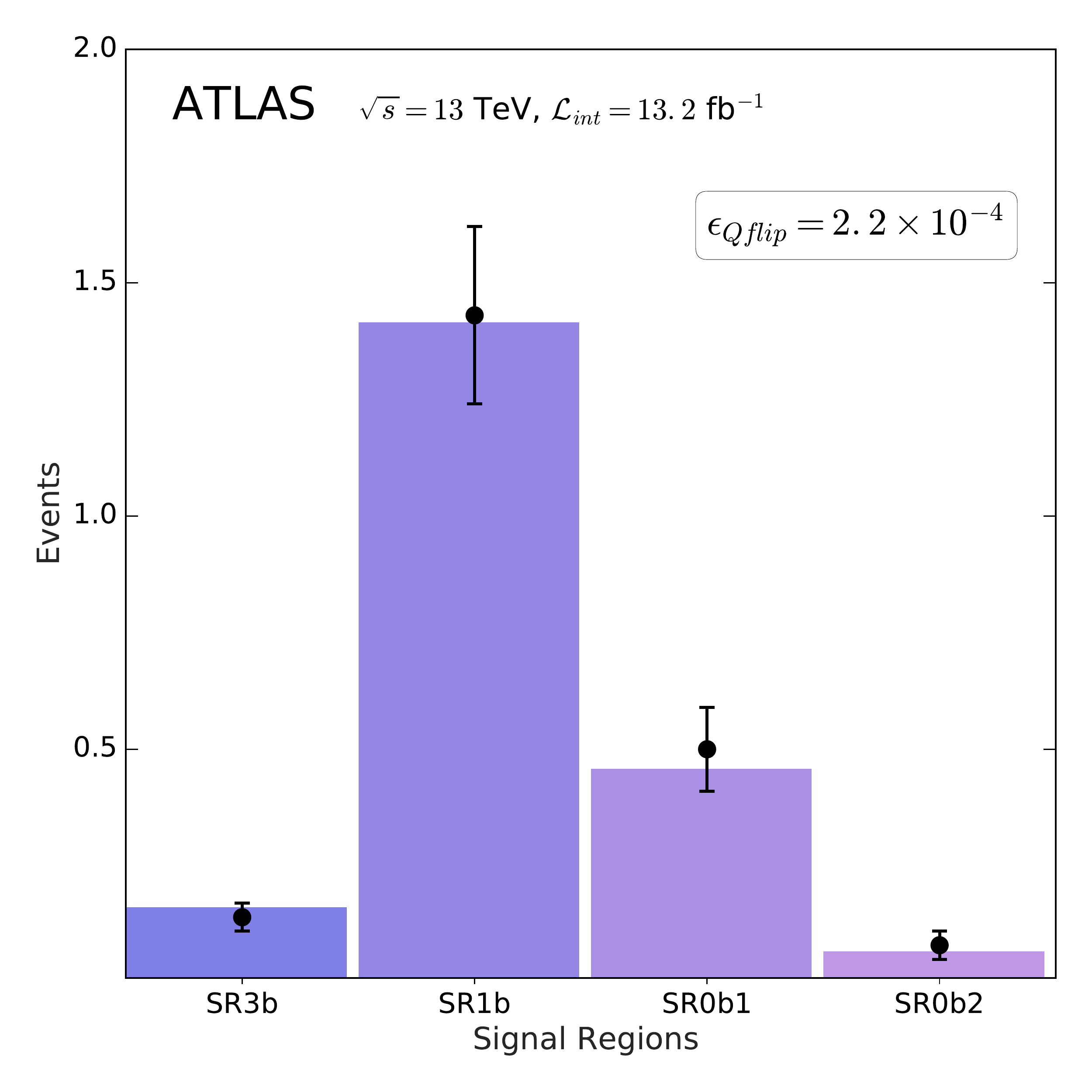}
\caption{\small Post-fit validation plots of our fake lepton and $Q$-flip background simulations (colored bars) for the best fit values given in Eq.~\eqref{fitted} compared to ATLAS data-driven estimations (black dots with error bars). The definitions of the signal regions are given in Ref.~\cite{ATLAS:2016kjm}. 
\small  }
\label{Fake-Qflip-validation}
\end{center}
\end{figure}

The usual way to estimate fake lepton and $Q$-flip backgrounds is by directly extracting the probability of object mis-identification as a function of $p_T$ and $\eta$ from control data samples in signal regions relevant to the proposed search. Electron $Q$-flip rates are estimated using a likelihood fit to $Z/\gamma^*\to ee$ data samples \cite{ATLAS:2016kjm}, while lepton fake rates are extracted using the matrix method applied to semi-leptonic $t\bar t$ data samples \cite{Aad:2016tuk}. The extracted values are typically in the range $\epsilon_{\text{fake}}\sim\mathcal{O}(10^{-4})$ and $\epsilon_{\text{Qflip}}\sim\mathcal{O}(10^{-3})$ or less. For this reason many phenomenological analysis adopt as benchmark values:
\begin{equation}\label{bench}
\epsilon_{\text{fake}}=10^{-4} \, , \ \ \ \ \ \ \ \ \ \ \ \epsilon_{\text{Qflip}}=10^{-3}\,.
\end{equation}
In order to better assess the sensitivity of our four-top search, it is important to know if this benchmark choice is either too conservative or too optimistic. The data-driven techniques used by the experimental collaborations are however difficult to implement without access to the 13 TeV control data sets. Nonetheless, we can still estimate $\epsilon_{\text{fake,Qflip}}$ by combining MC simulations and existing LHC searches as proposed in Ref.~\cite{Curtin:2013zua}. For this we rely on the SUSY search \cite{ATLAS:2016kjm} by ATLAS at 13 TeV and $13.2$ fb$^{-1}$, where SS dileptons and trileptons were used as final states in different signal regions similar to the one in our multi-lepton search (for example {\bf SR3b}). There, fake lepton and $Q$-flip backgrounds were estimated using data-driven techniques. We make use of their results (Tables~4 and 5 in Ref.~\cite{ATLAS:2016kjm}) to match  MC samples by fitting the parameters $\epsilon_{\text{fake,Qflip}}$ that scale the total number of events in each signal region. For the fit we used samples of $t\bar t$ + jets, $W^\pm$ + jets and $Z$ + jets followed by a $j\to\ell^\pm$ or $e^\pm\to e^\mp$ mis-identification. The best fit values we obtain are:
\begin{equation}\label{fitted}
\epsilon_{\text{fake}}=7.2\times10^{-5} \, , \ \ \ \ \ \ \epsilon_{\text{Qflip}}=2.2\times10^{-4}\,.
\end{equation}
Post-fit results of our fake lepton and $Q$-flip background simulations for this search, given by the colored bars in Fig.~\ref{Fake-Qflip-validation}, agree very well within error bars with the ATLAS data-driven estimations in each signal region (black dots). Our fit suggests that the benchmark probabilities in Eq.~\eqref{bench} are a conservative choice for our four-top search and that the number reducible backgrounds events presented in Table~\ref{ss-results} and Table.~\ref{trilep-results} between parenthesis are over-estimated. For this reason we present the results in Sec.~\ref{section:3.3} based on the estimation in Eq.~\eqref{fitted} instead. In principle, a better fit to the ATLAS results could be achieved by including $p_T$ and $\eta$ dependence for each probability and if more fitting parameters are included \cite{Curtin:2013zua}, but this should not alter considerably the values in Eq.~\eqref{fitted}.

\subsection{ Non-prompt lepton cuts}
\label{btag-NPL}

\begin{figure}h!]
\begin{center}
\includegraphics[width=0.6\textwidth]{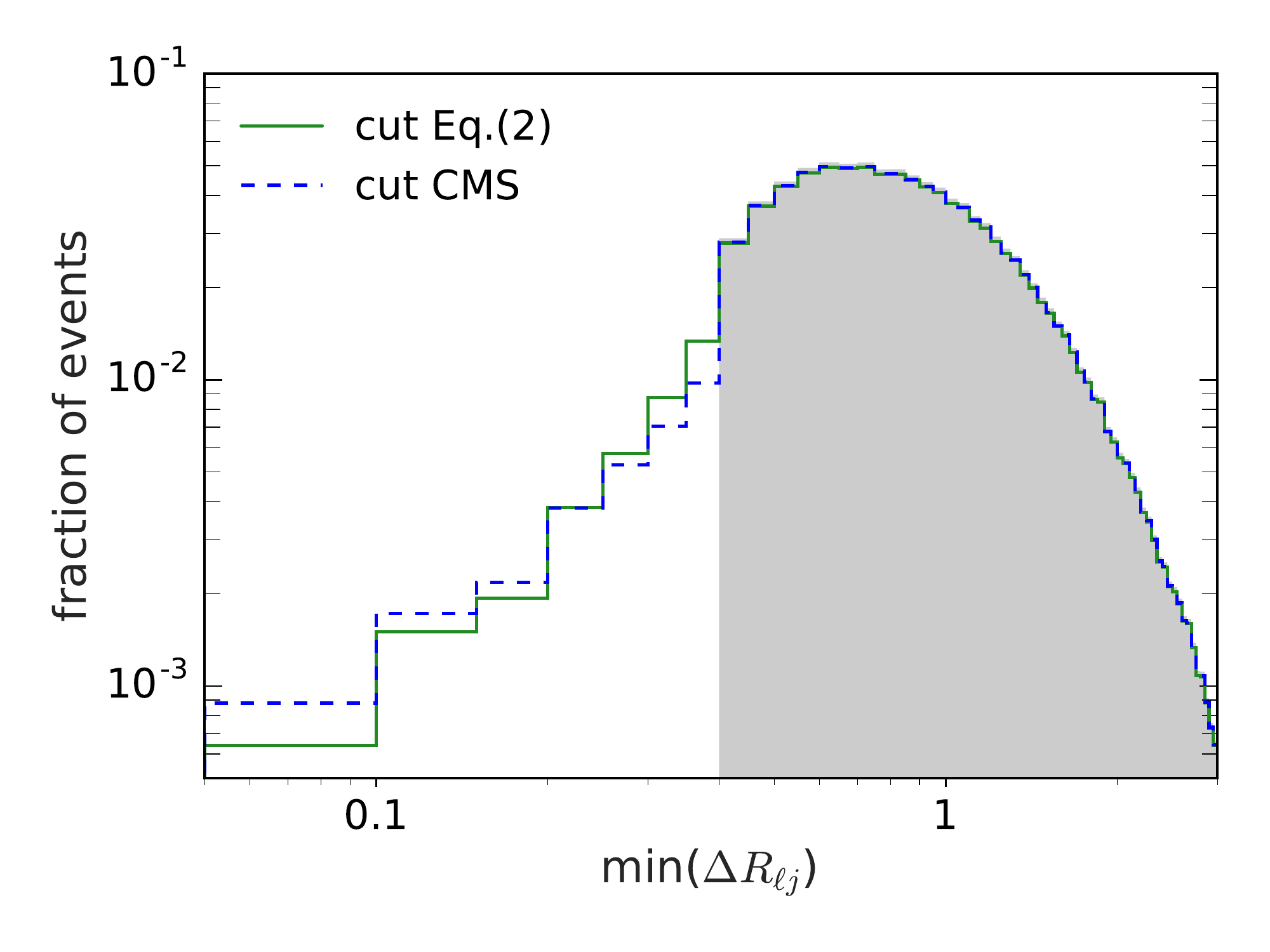}
\caption{ \small Comparison between different non-prompt lepton cuts for the four-top signal: $\Delta R_{\ell j}<0.4$ veto (shaded gray), the cut adopted by CMS in Ref.~\cite{CMS:2016vfu} (dashed blue) and the cut in Eq.~\eqref{nonprompt} adopted in our analysis (green).}
\label{btag_nonprompt} 
\end{center}
\end{figure}

Here we give a few comments on the cut in Eq.~\eqref{nonprompt} applied to leptons passing the mini-isolation requirement. Most analyses by ATLAS and CMS veto reconstructed leptons too close to jets (typically with a distance $\Delta R_{\ell j}<0.4$ from any jet) in order to reduce (isolated) non-prompt lepton backgrounds originating from heavy meson decays inside jets. A slightly looser cut has been adopted by CMS in Ref.~\cite{CMS:2016vfu}, based on rejecting isolated leptons satisfying:
\begin{equation}\label{CMScut}
 p_{T_\ell} < \alpha\ p_{T_{\text{jet}}}  
\end{equation}
\noindent with $\alpha=0.7$ for electrons and $\alpha=0.85$ for muons. With these values, CMS claims a high background rejection while keeping more than 50\% of the signal inside the $\Delta R_{\ell j}<0.4$ region. The non-prompt lepton cut in Eq.~\eqref{nonprompt} used in our search has an additional dependence on the distance of the lepton from the jet, making the cut tighter (looser) at closer (larger) distances than the one in Eq.~\eqref{CMScut}. In Fig.~\ref{btag_nonprompt} we show the fractions of four-top events after implementing the non-prompt lepton cuts: $\Delta R_{\ell j}<0.4$ veto (shaded gray), the cut adopted by CMS in Ref.~\cite{CMS:2016vfu} (dashed blue) and the cut in Eq.~\eqref{nonprompt} adopted in our analysis (green).

\bibliographystyle{JHEP}
\bibliography{biblio}

\providecommand{\href}[2]{#2}\begingroup\raggedright\begin{thebibliography}{10}

\bibitem{ATLAS:2016lgh}
{\scshape ATLAS} collaboration, T.~A. collaboration, \emph{{Search for Higgs
  boson production via weak boson fusion and decaying to $b \bar b$ in
  association with a high-energy photon in the ATLAS detector}},
  \href{http://arxiv.org/abs/ATLAS-CONF-2016-063}{{\tt ATLAS-CONF-2016-063}}.

\bibitem{Aad:2015tna}
{\scshape ATLAS} collaboration, G.~Aad et~al., \emph{{Constraints on
  non-Standard Model Higgs boson interactions in an effective Lagrangian using
  differential cross sections measured in the $H \rightarrow \gamma\gamma$
  decay channel at $\sqrt{s} = 8$TeV with the ATLAS detector}},
  \href{http://dx.doi.org/10.1016/j.physletb.2015.11.071}{\emph{Phys. Lett.}
  {\bf B753} (2016) 69--85}, [\href{http://arxiv.org/abs/1508.02507}{{\tt
  1508.02507}}].

\bibitem{Aad:2015lha}
{\scshape ATLAS} collaboration, G.~Aad et~al., \emph{{Measurements of the Total
  and Differential Higgs Boson Production Cross Sections Combining the $H \to
  \gamma \gamma$ and $H \to Z Z^{*} \to 4l$ Decay Channels at $\sqrt{s}$=8TeV
  with the ATLAS Detector}},
  \href{http://dx.doi.org/10.1103/PhysRevLett.115.091801}{\emph{Phys. Rev.
  Lett.} {\bf 115} (2015) 091801}, [\href{http://arxiv.org/abs/1504.05833}{{\tt
  1504.05833}}].

\bibitem{Khachatryan:2015yvw}
{\scshape CMS} collaboration, V.~Khachatryan et~al., \emph{{Measurement of
  differential and integrated fiducial cross sections for Higgs boson
  production in the four-lepton decay channel in pp collisions at $ \sqrt{s}=7
  $ and 8 TeV}}, \href{http://dx.doi.org/10.1007/JHEP04(2016)005}{\emph{JHEP}
  {\bf 04} (2016) 005}, [\href{http://arxiv.org/abs/1512.08377}{{\tt
  1512.08377}}].

\bibitem{Khachatryan:2015rxa}
{\scshape CMS} collaboration, V.~Khachatryan et~al., \emph{{Measurement of
  differential cross sections for Higgs boson production in the diphoton decay
  channel in pp collisions at $\sqrt{s}=8\,\text {TeV} $}},
  \href{http://dx.doi.org/10.1140/epjc/s10052-015-3853-3}{\emph{Eur. Phys. J.}
  {\bf C76} (2016) 13}, [\href{http://arxiv.org/abs/1508.07819}{{\tt
  1508.07819}}].

\bibitem{CMS:2016cdj}
{\scshape CMS} collaboration, C.~Collaboration, \emph{{Search for Higgs boson
  pair production in the $\mathrm{b}\overline{\mathrm{b}} \mathrm{l}\nu
  \mathrm{l}\nu$ final state at $\sqrt{s} = 13~\mathrm{TeV}$}},
  \href{http://arxiv.org/abs/CMS-PAS-HIG-16-024}{{\tt CMS-PAS-HIG-16-024}}.

\bibitem{Aaboud:2016xco}
{\scshape ATLAS} collaboration, M.~Aaboud et~al., \emph{{Search for pair
  production of Higgs bosons in the $b\bar{b}b\bar{b}$ final state using
  proton--proton collisions at $\sqrt{s} = 13$ TeV with the ATLAS detector}},
  \href{http://dx.doi.org/10.1103/PhysRevD.94.052002}{\emph{Phys. Rev.} {\bf
  D94} (2016) 052002}, [\href{http://arxiv.org/abs/1606.04782}{{\tt
  1606.04782}}].

\bibitem{Khachatryan:2016sey}
{\scshape CMS} collaboration, V.~Khachatryan et~al., \emph{{Search for two
  Higgs bosons in final states containing two photons and two bottom quarks}},
  {\emph{Submitted to: Phys. Rev. D} (2016) },
  [\href{http://arxiv.org/abs/1603.06896}{{\tt 1603.06896}}].

\bibitem{ATLAS:2016pkl}
{\scshape ATLAS} collaboration, T.~A. collaboration, \emph{{Search for the
  Standard Model Higgs boson produced in association with a vector boson and
  decaying to a $b\bar{b}$ pair in $pp$ collisions at 13 TeV using the ATLAS
  detector}},  \href{http://arxiv.org/abs/ATLAS-CONF-2016-091}{{\tt
  ATLAS-CONF-2016-091}}.

\bibitem{CMS:2016ygt}
{\scshape CMS} collaboration, C.~Collaboration, \emph{{Search for H to bbar in
  association with a single top quark as a test of Higgs boson couplings at 13
  TeV}},  \href{http://arxiv.org/abs/CMS-PAS-HIG-16-019}{{\tt
  CMS-PAS-HIG-16-019}}.

\bibitem{CMS:2016vqb}
{\scshape CMS} collaboration, C.~Collaboration, \emph{{Search for associated
  production of Higgs bosons and top quarks in multilepton final states at
  $\sqrt{s}=13~\mathrm{TeV}$}},
  \href{http://arxiv.org/abs/CMS-PAS-HIG-16-022}{{\tt CMS-PAS-HIG-16-022}}.

\bibitem{ATLAS:2016axz}
{\scshape ATLAS} collaboration, T.~A. collaboration, \emph{{Combination of the
  searches for Higgs boson production in association with top quarks in the
  $\gamma\gamma$, multilepton, and $b\bar{b}$ decay channels at $\sqrt{s}$=13
  TeV with the ATLAS Detector}},
  \href{http://arxiv.org/abs/ATLAS-CONF-2016-068}{{\tt ATLAS-CONF-2016-068}}.

\bibitem{ATLAS:2016soq}
{\scshape ATLAS} collaboration, T.~A. collaboration, \emph{{Measurements of
  top-quark pair differential cross-sections in the lepton+jets channel in pp
  collisions at $\sqrt{s}=13$ TeV using the ATLAS detector}},
  \href{http://arxiv.org/abs/ATLAS-CONF-2016-040}{{\tt ATLAS-CONF-2016-040}}.

\bibitem{CMS:2016xyh}
{\scshape CMS} collaboration, C.~Collaboration, \emph{{Measurement of particle
  level differential ttbar cross sections in the dilepton channel at sqrt(s) =
  13 TeV}},  \href{http://arxiv.org/abs/CMS-PAS-TOP-16-007}{{\tt
  CMS-PAS-TOP-16-007}}.

\bibitem{Aaboud:2016iot}
{\scshape ATLAS} collaboration, M.~Aaboud et~al., \emph{{Measurement of top
  quark pair differential cross-sections in the dilepton channel in $pp$
  collisions at $\sqrt{s}$ = 7 and 8 TeV with ATLAS}},
  \href{http://arxiv.org/abs/1607.07281}{{\tt 1607.07281}}.

\bibitem{CMS:2016cue}
{\scshape CMS} collaboration, C.~Collaboration, \emph{{Measurement of double
  differential cross sections for top quark pair production in pp collisions at
  sqrt[s] = 8 TeV}},  \href{http://arxiv.org/abs/CMS-PAS-TOP-14-013}{{\tt
  CMS-PAS-TOP-14-013}}.

\bibitem{ATLAS:2016lte}
{\scshape ATLAS} collaboration, T.~A. collaboration, \emph{{Measurement of the
  cross-section of the production of a $W$ boson in association with a single
  top quark with ATLAS at $\sqrt{s}=13\,\mathrm{TeV}$}},
  \href{http://arxiv.org/abs/ATLAS-CONF-2016-065}{{\tt ATLAS-CONF-2016-065}}.

\bibitem{CMS:2016ufa}
{\scshape CMS} collaboration, C.~Collaboration, \emph{{Combination of
  cross-section measurements of associated production of a single top quark and
  a W boson at $\sqrt{s} = 8~\mathrm{TeV}$ with the ATLAS and CMS
  experiments}},  \href{http://arxiv.org/abs/CMS-PAS-TOP-15-019}{{\tt
  CMS-PAS-TOP-15-019}}.

\bibitem{ATLAS-CONF-2016-003}
\emph{{Measurement of the $t\bar{t}Z$ and $t\bar{t}W$ production cross sections
  in multilepton final states using 3.2 fb$^{-1}$ of $pp$ collisions at 13 TeV
  at the LHC}},  Tech. Rep. ATLAS-CONF-2016-003, CERN, Geneva, Mar, 2016.

\bibitem{CMS:2016dui}
{\scshape CMS} collaboration, C.~Collaboration, \emph{{Measurement of the top
  pair-production in association with a W or Z boson in pp collisions at 13
  TeV}},  \href{http://arxiv.org/abs/CMS-PAS-TOP-16-017}{{\tt
  CMS-PAS-TOP-16-017}}.

\bibitem{Lillie:2007hd}
B.~Lillie, J.~Shu and T.~M.~P. Tait, \emph{{Top Compositeness at the Tevatron
  and LHC}}, \href{http://dx.doi.org/10.1088/1126-6708/2008/04/087}{\emph{JHEP}
  {\bf 04} (2008) 087}, [\href{http://arxiv.org/abs/0712.3057}{{\tt
  0712.3057}}].

\bibitem{Pomarol:2008bh}
A.~Pomarol and J.~Serra, \emph{{Top Quark Compositeness: Feasibility and
  Implications}},
  \href{http://dx.doi.org/10.1103/PhysRevD.78.074026}{\emph{Phys. Rev.} {\bf
  D78} (2008) 074026}, [\href{http://arxiv.org/abs/0806.3247}{{\tt
  0806.3247}}].

\bibitem{Kumar:2009vs}
K.~Kumar, T.~M.~P. Tait and R.~Vega-Morales, \emph{{Manifestations of Top
  Compositeness at Colliders}},
  \href{http://dx.doi.org/10.1088/1126-6708/2009/05/022}{\emph{JHEP} {\bf 05}
  (2009) 022}, [\href{http://arxiv.org/abs/0901.3808}{{\tt 0901.3808}}].

\bibitem{Cacciapaglia:2011kz}
G.~Cacciapaglia, R.~Chierici, A.~Deandrea, L.~Panizzi, S.~Perries and S.~Tosi,
  \emph{{Four tops on the real projective plane at LHC}},
  \href{http://dx.doi.org/10.1007/JHEP10(2011)042}{\emph{JHEP} {\bf 10} (2011)
  042}, [\href{http://arxiv.org/abs/1107.4616}{{\tt 1107.4616}}].

\bibitem{Perelstein:2011ez}
M.~Perelstein and A.~Spray, \emph{{Four boosted tops from a Regge gluon}},
  \href{http://dx.doi.org/10.1007/JHEP09(2011)008}{\emph{JHEP} {\bf 09} (2011)
  008}, [\href{http://arxiv.org/abs/1106.2171}{{\tt 1106.2171}}].

\bibitem{AguilarSaavedra:2011ck}
J.~A. Aguilar-Saavedra and J.~Santiago, \emph{{Four tops and the $t \bar{t}$
  forward-backward asymmetry}},
  \href{http://dx.doi.org/10.1103/PhysRevD.85.034021}{\emph{Phys. Rev.} {\bf
  D85} (2012) 034021}, [\href{http://arxiv.org/abs/1112.3778}{{\tt
  1112.3778}}].

\bibitem{Beck:2015cga}
L.~Beck, F.~Blekman, D.~Dobur, B.~Fuks, J.~Keaveney and K.~Mawatari,
  \emph{{Probing top-philic sgluons with LHC Run I data}},
  \href{http://dx.doi.org/10.1016/j.physletb.2015.04.043}{\emph{Phys. Lett.}
  {\bf B746} (2015) 48--52}, [\href{http://arxiv.org/abs/1501.07580}{{\tt
  1501.07580}}].

\bibitem{Dev:2014yca}
P.~S. Bhupal~Dev and A.~Pilaftsis, \emph{{Maximally Symmetric Two Higgs Doublet
  Model with Natural Standard Model Alignment}},
  \href{http://dx.doi.org/10.1007/JHEP11(2015)147,
  10.1007/JHEP12(2014)024}{\emph{JHEP} {\bf 12} (2014) 024},
  [\href{http://arxiv.org/abs/1408.3405}{{\tt 1408.3405}}].

\bibitem{Bevilacqua:2012em}
G.~Bevilacqua and M.~Worek, \emph{{Constraining BSM Physics at the LHC: Four
  top final states with NLO accuracy in perturbative QCD}},
  \href{http://dx.doi.org/10.1007/JHEP07(2012)111}{\emph{JHEP} {\bf 07} (2012)
  111}, [\href{http://arxiv.org/abs/1206.3064}{{\tt 1206.3064}}].

\bibitem{Gerbush:2007fe}
M.~Gerbush, T.~J. Khoo, D.~J. Phalen, A.~Pierce and D.~Tucker-Smith,
  \emph{{Color-octet scalars at the CERN LHC}},
  \href{http://dx.doi.org/10.1103/PhysRevD.77.095003}{\emph{Phys. Rev.} {\bf
  D77} (2008) 095003}, [\href{http://arxiv.org/abs/0710.3133}{{\tt
  0710.3133}}].

\bibitem{Acharya:2009gb}
B.~S. Acharya, P.~Grajek, G.~L. Kane, E.~Kuflik, K.~Suruliz and L.-T. Wang,
  \emph{{Identifying Multi-Top Events from Gluino Decay at the LHC}},
  \href{http://arxiv.org/abs/0901.3367}{{\tt 0901.3367}}.

\bibitem{Gregoire:2011ka}
T.~Gregoire, E.~Katz and V.~Sanz, \emph{{Four top quarks in extensions of the
  standard model}},
  \href{http://dx.doi.org/10.1103/PhysRevD.85.055024}{\emph{Phys. Rev.} {\bf
  D85} (2012) 055024}, [\href{http://arxiv.org/abs/1101.1294}{{\tt
  1101.1294}}].

\bibitem{Liu:2015hxi}
D.~Liu and R.~Mahbubani, \emph{{Probing top-antitop resonances with $t\bar{t}$
  scattering at LHC14}},
  \href{http://dx.doi.org/10.1007/JHEP04(2016)116}{\emph{JHEP} {\bf 04} (2016)
  116}, [\href{http://arxiv.org/abs/1511.09452}{{\tt 1511.09452}}].

\bibitem{Gori:2016zto}
S.~Gori, I.-W. Kim, N.~R. Shah and K.~M. Zurek, \emph{{Closing the Wedge:
  Search Strategies for Extended Higgs Sectors with Heavy Flavor Final
  States}}, \href{http://dx.doi.org/10.1103/PhysRevD.93.075038}{\emph{Phys.
  Rev.} {\bf D93} (2016) 075038}, [\href{http://arxiv.org/abs/1602.02782}{{\tt
  1602.02782}}].

\bibitem{Kim:2016plm}
J.~H. Kim, K.~Kong, S.~J. Lee and G.~Mohlabeng, \emph{{Probing TeV scale
  Top-Philic Resonances with Boosted Top-Tagging at the High Luminosity LHC}},
  \href{http://dx.doi.org/10.1103/PhysRevD.94.035023}{\emph{Phys. Rev.} {\bf
  D94} (2016) 035023}, [\href{http://arxiv.org/abs/1604.07421}{{\tt
  1604.07421}}].

\bibitem{Khachatryan:2014sca}
{\scshape CMS} collaboration, V.~Khachatryan et~al., \emph{{Search for Standard
  Model Production of Four Top Quarks in the Lepton + Jets Channel in pp
  Collisions at $\sqrt{s}$ = 8 TeV}},
  \href{http://dx.doi.org/10.1007/JHEP11(2014)154}{\emph{JHEP} {\bf 11} (2014)
  154}, [\href{http://arxiv.org/abs/1409.7339}{{\tt 1409.7339}}].

\bibitem{Aad:2015kqa}
{\scshape ATLAS} collaboration, G.~Aad et~al., \emph{{Search for production of
  vector-like quark pairs and of four top quarks in the lepton-plus-jets final
  state in $pp$ collisions at $\sqrt{s}=8$ TeV with the ATLAS detector}},
  \href{http://dx.doi.org/10.1007/JHEP08(2015)105}{\emph{JHEP} {\bf 08} (2015)
  105}, [\href{http://arxiv.org/abs/1505.04306}{{\tt 1505.04306}}].

\bibitem{ATLAS:2016gqb}
{\scshape ATLAS} collaboration, T.~A. collaboration, \emph{{Search for
  four-top-quark production in final states with one charged lepton and
  multiple jets using 3.2 fb$^{-1}$ of proton-proton collisions at $\sqrt{s}$ =
  13 TeV with the ATLAS detector at the LHC}},
  \href{http://arxiv.org/abs/ATLAS-CONF-2016-020}{{\tt ATLAS-CONF-2016-020}}.

\bibitem{CMS:2016wig}
{\scshape CMS} collaboration, C.~Collaboration, \emph{{Search for standard
  model production of four top quarks in proton-proton collisions at 13 TeV}},
  \href{http://arxiv.org/abs/CMS-PAS-TOP-16-016}{{\tt CMS-PAS-TOP-16-016}}.

\bibitem{ATLAS:2016sno}
{\scshape ATLAS} collaboration, T.~A. collaboration, \emph{{Search for new
  physics using events with $b$-jets and a pair of same charge leptons in 3.2
  fb$^{-1}$ of $pp$ collisions at $\sqrt{s}=13$ TeV with the ATLAS detector}},
  \href{http://arxiv.org/abs/ATLAS-CONF-2016-032}{{\tt ATLAS-CONF-2016-032}}.

\bibitem{Aad:2016tuk}
{\scshape ATLAS} collaboration, G.~Aad et~al., \emph{{Search for supersymmetry
  at $\sqrt{s}=13$ TeV in final states with jets and two same-sign leptons or
  three leptons with the ATLAS detector}},
  \href{http://dx.doi.org/10.1140/epjc/s10052-016-4095-8}{\emph{Eur. Phys. J.}
  {\bf C76} (2016) 259}, [\href{http://arxiv.org/abs/1602.09058}{{\tt
  1602.09058}}].

\bibitem{CMS:2016ahn}
{\scshape CMS} collaboration, C.~Collaboration, \emph{{Search for SUSY with
  multileptons in 13 TeV data}},
  \href{http://arxiv.org/abs/CMS-PAS-SUS-16-022}{{\tt CMS-PAS-SUS-16-022}}.

\bibitem{Alwall:2014hca}
J.~Alwall, R.~Frederix, S.~Frixione, V.~Hirschi, F.~Maltoni, O.~Mattelaer
  et~al., \emph{{The automated computation of tree-level and next-to-leading
  order differential cross sections, and their matching to parton shower
  simulations}}, \href{http://dx.doi.org/10.1007/JHEP07(2014)079}{\emph{JHEP}
  {\bf 07} (2014) 079}, [\href{http://arxiv.org/abs/1405.0301}{{\tt
  1405.0301}}].

\bibitem{Thaler:2008ju}
J.~Thaler and L.-T. Wang, \emph{{Strategies to Identify Boosted Tops}},
  \href{http://dx.doi.org/10.1088/1126-6708/2008/07/092}{\emph{JHEP} {\bf 07}
  (2008) 092}, [\href{http://arxiv.org/abs/0806.0023}{{\tt 0806.0023}}].

\bibitem{Rehermann:2010vq}
K.~Rehermann and B.~Tweedie, \emph{{Efficient Identification of Boosted
  Semileptonic Top Quarks at the LHC}},
  \href{http://dx.doi.org/10.1007/JHEP03(2011)059}{\emph{JHEP} {\bf 03} (2011)
  059}, [\href{http://arxiv.org/abs/1007.2221}{{\tt 1007.2221}}].

\bibitem{Sjostrand:2014zea}
T.~Sjostrand, S.~Ask, J.~R. Christiansen, R.~Corke, N.~Desai, P.~Ilten et~al.,
  \emph{{An Introduction to PYTHIA 8.2}},
  \href{http://dx.doi.org/10.1016/j.cpc.2015.01.024}{\emph{Comput. Phys.
  Commun.} {\bf 191} (2015) 159--177},
  [\href{http://arxiv.org/abs/1410.3012}{{\tt 1410.3012}}].

\bibitem{deFavereau:2013fsa}
{\scshape DELPHES 3} collaboration, J.~de~Favereau, C.~Delaere, P.~Demin,
  A.~Giammanco, V.~Lemaitre, A.~Mertens et~al., \emph{{DELPHES 3, A modular
  framework for fast simulation of a generic collider experiment}},
  \href{http://dx.doi.org/10.1007/JHEP02(2014)057}{\emph{JHEP} {\bf 02} (2014)
  057}, [\href{http://arxiv.org/abs/1307.6346}{{\tt 1307.6346}}].

\bibitem{Zhou:2012dz}
N.~Zhou, D.~Whiteson and T.~M.~P. Tait, \emph{{Limits on Four-Top Production
  from the ATLAS Same-sign Top-quark Search}},
  \href{http://dx.doi.org/10.1103/PhysRevD.85.091501}{\emph{Phys. Rev.} {\bf
  D85} (2012) 091501}, [\href{http://arxiv.org/abs/1203.5862}{{\tt
  1203.5862}}].

\bibitem{CMS:2016vfu}
{\scshape CMS} collaboration, C.~Collaboration, \emph{{Search for SUSY in
  same-sign dilepton events at 13 TeV}},
  \href{http://arxiv.org/abs/CMS-PAS-SUS-16-020}{{\tt CMS-PAS-SUS-16-020}}.

\bibitem{Mangano:2002ea}
M.~L. Mangano, M.~Moretti, F.~Piccinini, R.~Pittau and A.~D. Polosa,
  \emph{{ALPGEN, a generator for hard multiparton processes in hadronic
  collisions}},
  \href{http://dx.doi.org/10.1088/1126-6708/2003/07/001}{\emph{JHEP} {\bf 07}
  (2003) 001}, [\href{http://arxiv.org/abs/hep-ph/0206293}{{\tt
  hep-ph/0206293}}].

\bibitem{ATL-PHYS-PUB-2015-022}
\emph{{Expected performance of the ATLAS $b$-tagging algorithms in Run-2}},
  Tech. Rep. ATL-PHYS-PUB-2015-022, CERN, Geneva, Jul, 2015.

\bibitem{Abazov:2013eha}
{\scshape D0} collaboration, V.~M. Abazov et~al., \emph{{Search for Higgs boson
  production in trilepton and like-charge electron-muon final states with the
  D0 detector}},
  \href{http://dx.doi.org/10.1103/PhysRevD.88.052009}{\emph{Phys. Rev.} {\bf
  D88} (2013) 052009}, [\href{http://arxiv.org/abs/1302.5723}{{\tt
  1302.5723}}].

\bibitem{Aaltonen:2013vca}
{\scshape CDF} collaboration, T.~A. Aaltonen et~al., \emph{{Search for new
  physics in trilepton events and limits on the associated chargino-neutralino
  production at CDF}},
  \href{http://dx.doi.org/10.1103/PhysRevD.90.012011}{\emph{Phys. Rev.} {\bf
  D90} (2014) 012011}, [\href{http://arxiv.org/abs/1309.7509}{{\tt
  1309.7509}}].

\bibitem{Cowan:2010js}
G.~Cowan, K.~Cranmer, E.~Gross and O.~Vitells, \emph{{Asymptotic formulae for
  likelihood-based tests of new physics}},
  \href{http://dx.doi.org/10.1140/epjc/s10052-011-1554-0,
  10.1140/epjc/s10052-013-2501-z}{\emph{Eur. Phys. J.} {\bf C71} (2011) 1554},
  [\href{http://arxiv.org/abs/1007.1727}{{\tt 1007.1727}}].

\bibitem{ATLAS:2016kjm}
{\scshape ATLAS} collaboration, T.~A. collaboration, \emph{{Search for
  supersymmetry with two same-sign leptons or three leptons using 13.2
  fb$^{-1}$ of $\sqrt{s} = 13$ TeV $pp$ collision data collected by the ATLAS
  detector}},  \href{http://arxiv.org/abs/ATLAS-CONF-2016-037}{{\tt
  ATLAS-CONF-2016-037}}.

\bibitem{Kane:2011zd}
G.~L. Kane, E.~Kuflik, R.~Lu and L.-T. Wang, \emph{{Top Channel for Early SUSY
  Discovery at the LHC}},
  \href{http://dx.doi.org/10.1103/PhysRevD.84.095004}{\emph{Phys. Rev.} {\bf
  D84} (2011) 095004}, [\href{http://arxiv.org/abs/1101.1963}{{\tt
  1101.1963}}].

\bibitem{Haisch:2016hzu}
U.~Haisch and J.~F. Kamenik, \emph{{Searching for new spin-0 resonances at
  LHCb}}, \href{http://dx.doi.org/10.1103/PhysRevD.93.055047}{\emph{Phys. Rev.}
  {\bf D93} (2016) 055047}, [\href{http://arxiv.org/abs/1601.05110}{{\tt
  1601.05110}}].

\bibitem{Fajfer:2012jt}
S.~Fajfer, J.~F. Kamenik, I.~Nisandzic and J.~Zupan, \emph{{Implications of
  Lepton Flavor Universality Violations in B Decays}},
  \href{http://dx.doi.org/10.1103/PhysRevLett.109.161801}{\emph{Phys. Rev.
  Lett.} {\bf 109} (2012) 161801}, [\href{http://arxiv.org/abs/1206.1872}{{\tt
  1206.1872}}].

\bibitem{Faroughy:2016osc}
D.~A. Faroughy, A.~Greljo and J.~F. Kamenik, \emph{{Confronting lepton flavor
  universality violation in B decays with high-$p_T$ tau lepton searches at
  LHC}},  \href{http://arxiv.org/abs/1609.07138}{{\tt 1609.07138}}.

\bibitem{Dolan:2014ska}
M.~J. Dolan, F.~Kahlhoefer, C.~McCabe and K.~Schmidt-Hoberg, \emph{{A taste of
  dark matter: Flavour constraints on pseudoscalar mediators}},
  \href{http://dx.doi.org/10.1007/JHEP07(2015)103,
  10.1007/JHEP03(2015)171}{\emph{JHEP} {\bf 03} (2015) 171},
  [\href{http://arxiv.org/abs/1412.5174}{{\tt 1412.5174}}].

\bibitem{Arina:2016cqj}
C.~Arina et~al., \emph{{A comprehensive approach to dark matter studies:
  exploration of simplified top-philic models}},
  \href{http://arxiv.org/abs/1605.09242}{{\tt 1605.09242}}.

\bibitem{Jackson:2009kg}
C.~B. Jackson, G.~Servant, G.~Shaughnessy, T.~M.~P. Tait and M.~Taoso,
  \emph{{Higgs in Space!}},
  \href{http://dx.doi.org/10.1088/1475-7516/2010/04/004}{\emph{JCAP} {\bf 1004}
  (2010) 004}, [\href{http://arxiv.org/abs/0912.0004}{{\tt 0912.0004}}].

\bibitem{Jackson:2013rqp}
C.~B. Jackson, G.~Servant, G.~Shaughnessy, T.~M.~P. Tait and M.~Taoso,
  \emph{{Gamma Rays from Top-Mediated Dark Matter Annihilations}},
  \href{http://dx.doi.org/10.1088/1475-7516/2013/07/006}{\emph{JCAP} {\bf 1307}
  (2013) 006}, [\href{http://arxiv.org/abs/1303.4717}{{\tt 1303.4717}}].

\bibitem{Goertz:2015nkp}
F.~Goertz, J.~F. Kamenik, A.~Katz and M.~Nardecchia, \emph{{Indirect
  Constraints on the Scalar Di-Photon Resonance at the LHC}},
  \href{http://dx.doi.org/10.1007/JHEP05(2016)187}{\emph{JHEP} {\bf 05} (2016)
  187}, [\href{http://arxiv.org/abs/1512.08500}{{\tt 1512.08500}}].

\bibitem{Kamenik:2011vy}
J.~F. Kamenik and C.~Smith, \emph{{FCNC portals to the dark sector}},
  \href{http://dx.doi.org/10.1007/JHEP03(2012)090}{\emph{JHEP} {\bf 03} (2012)
  090}, [\href{http://arxiv.org/abs/1111.6402}{{\tt 1111.6402}}].

\bibitem{Dicus:1994bm}
D.~Dicus, A.~Stange and S.~Willenbrock, \emph{{Higgs decay to top quarks at
  hadron colliders}},
  \href{http://dx.doi.org/10.1016/0370-2693(94)91017-0}{\emph{Phys. Lett.} {\bf
  B333} (1994) 126--131}, [\href{http://arxiv.org/abs/hep-ph/9404359}{{\tt
  hep-ph/9404359}}].

\bibitem{Carena:2016npr}
M.~Carena and Z.~Liu, \emph{{Challenges and opportunities for heavy scalar
  searches in the $t\bar t$ channel at the LHC}},
  \href{http://arxiv.org/abs/1608.07282}{{\tt 1608.07282}}.

\bibitem{Craig:2016ygr}
N.~Craig, J.~Hajer, Y.-Y. Li, T.~Liu and H.~Zhang, \emph{{Heavy Higgs Bosons at
  Low $\tan \beta$: from the LHC to 100 TeV}},
  \href{http://arxiv.org/abs/1605.08744}{{\tt 1605.08744}}.

\bibitem{Cacciari:2011ma}
G.~S. M.Cacciari, G. P.~Salam, \emph{{FastJet user manual}},
  \href{http://dx.doi.org/10.1140/epjc/s10052-012-1896-2}{\emph{Eur. Phys. J.}
  {\bf C72} (2012) 1896}, [\href{http://arxiv.org/abs/1111.6097}{{\tt
  1111.6097}}].

\bibitem{Cacciari:2005hq}
M.~Cacciari and G.~P. Salam, \emph{{Dispelling the $N^{3}$ myth for the $k_t$
  jet-finder}},
  \href{http://dx.doi.org/10.1140/epjc/s10052-012-1896-2}{\emph{Phys. Lett.}
  {\bf B641} (2006) 57--61}, [\href{http://arxiv.org/abs/hep-ph/0512210}{{\tt
  hep-ph/0512210}}].

\bibitem{Curtin:2013zua}
D.~Curtin, J.~Galloway and J.~G. Wacker, \emph{{Measuring the $t \bar th$
  coupling from same-sign dilepton $+2b$ measurements}},
  \href{http://dx.doi.org/10.1103/PhysRevD.88.093006}{\emph{Phys. Rev.} {\bf
  D88} (2013) 093006}, [\href{http://arxiv.org/abs/1306.5695}{{\tt
  1306.5695}}].

\end{thebibliography}\endgroup
\end{document}